\documentclass[prd,preprint,superscriptaddress,showpacs,nofootinbib,tightenlines,amsmath]{revtex4}
\usepackage{graphics}
\usepackage{epsfig}
\usepackage{slashed}
\usepackage{pbox, calc}
\usepackage{braket}
\usepackage{nicefrac}
\usepackage{bbm}
\usepackage{amssymb}
\usepackage{gensymb}
\usepackage{bbold}
\usepackage{color}
\begin{document}
\allowdisplaybreaks
\title{Quark-mass and $1/N_c$ corrections to the vector-meson
pseudoscalar-meson photon ($VP\gamma$) interaction}
\author{H.~C.~Lange}
\affiliation{Institut f\"ur Kernphysik, Johannes Gutenberg-Universit\"at Mainz, D-55099 Mainz, Germany}
\author{A.~Krasniqi}
\affiliation{Institut f\"ur Kernphysik, Johannes Gutenberg-Universit\"at Mainz, D-55099 Mainz, Germany}
\author{S.~Scherer}
\affiliation{Institut f\"ur Kernphysik, Johannes Gutenberg-Universit\"at Mainz, D-55099 Mainz, Germany}
\date{November 12, 2021}
\begin{abstract}
   We analyze quark-mass and $1/N_c$ corrections to all of the radiative 
transitions between the vector-meson nonet and the pseudoscalar-meson nonet within a chiral effective Lagrangian approach.
     We perform fits of the available coupling constants to experimental data 
and discuss the corresponding approximations.
   In terms of five (six) coupling constants, we obtain a reasonably good
description of the 12 experimental decay rates.
\end{abstract}
\maketitle

\section{Introduction}

   Because of chiral symmetry and its spontaneous symmetry breaking in the 
ground state of quantum chromodynamics (QCD) \cite{Gasser:1982ap},
the members of the lowest-lying pseudoscalar octet ($\pi, K,\eta_8$) play a special role: they are the Goldstone bosons \cite{Goldstone:1961eq,Goldstone:1962es} of QCD and would be exactly massless for massless quarks.
   In the large-number-of-colors (large-$N_c$) limit
\cite{'tHooft:1973jz,Witten:1979kh}, i.e., $N_c\to\infty$ with $g^2N_c$ fixed, also the singlet eta, $\eta_1$, would be a Goldstone boson and would combine with the octet into a nonet of massless Goldstone bosons \cite{DiVecchia:1980yfw,Coleman:1980mx}.
   In the real world with $N_c=3$, the masses of the light pseudoscalars 
originate from an explicit symmetry breaking due to the quark masses \cite{Gasser:1982ap} and from the anomaly \cite{Adler:1969gk,Bell:1969ts} of the singlet axial-vector current \cite{'tHooft:1976up,Witten:1979vv,Veneziano:1979ec}.
   Chiral perturbation theory (ChPT) 
\cite{Weinberg:1978kz,Gasser:1983yg,Gasser:1984gg} provides a
systematic method of analyzing the low-energy interactions of the octet Goldstone bosons among each other and with external sources
(see, e.g., Refs.~\cite{Donoghue:1992dd,Scherer:2002tk,Scherer:2012zzd} for an 
introduction).
   The dynamical variables of ChPT are the Goldstone bosons rather than the 
quarks and gluons of QCD.
   By considering the combined chiral and large-$N_c$ limits, it is possible to 
set up Large-$N_c$ ChPT as the effective field theory of QCD at low energies including the singlet field~\cite{Moussallam:1994xp,Leutwyler:1996sa,
HerreraSiklody:1996pm,Leutwyler:1997yr,Kaiser:1998ds,HerreraSiklody:1998cr,
Kaiser:2000gs,Borasoy:2004ua,Guo:2015xva,Bickert:2016fgy}.

   Chiral symmetry also constrains the interactions of Goldstone bosons with 
heavier, i.e., non-Goldstone-boson hadrons, however, setting up a consistent power-counting scheme turns out to be more complex (see, e.g., Refs.~\cite{Gasser:1987rb,Jenkins:1990jv,Bijnens:1997rv,Becher:1999he,
Gegelia:1999gf,Fuchs:2003qc,Bruns:2004tj,Lutz:2008km,Djukanovic:2009zn}).
   Ever since the pioneering works on nonlinear realizations of chiral symmetry
\cite{Weinberg:1968de,Coleman:1969sm,Callan:1969sn},
there have been numerous approaches to the construction of chiral effective 
Lagrangians including vector mesons (see, e.g., Refs.~\cite{Gasser:1983yg,Kaymakcalan:1983qq,Kaymakcalan:1984bz,Bando:1985rf,
Meissner:1987ge,Bando:1987br,Ecker:1988te,Ecker:1989yg,
Birse:1996hd,Harada:2003jx,Kampf:2006yf,Djukanovic:2010tb}).
   They differ by, firstly, how the Lorentz group acts on the dynamical fields 
representing the vector mesons, either in terms of a vector field $V^\mu$ \cite{Weinberg:1995mt,Ryder:1985wq}
or in terms of an antisymmetric second-rank tensor field $T^{\mu\nu}$ \cite{Kyriakopoulos:1969zm,Kyriakopoulos:1973pt}, and, secondly, how the chiral group operates on the SU(3) flavor degrees of freedom of the vector mesons.
   The vector-meson pseudoscalar-meson photon ($VP\gamma$) interaction 
responsible for, e.g., the radiative decay of a vector meson
into a pseudoscalar meson is complementary to the hadronic decay of a vector meson into two pseudoscalar mesons, because it probes the so-called
odd-intrinsic-parity sector of low-energy QCD.
   In the present case, this refers to the odd number of Goldstone bosons, 
namely, one, participating in the interaction with a single vector meson and a photon.
   Starting with the early predictions based on SU(3) symmetry
\cite{Glashow:1963zz}, radiative decays of vector mesons into pseudoscalar
mesons were studied in a large number of approaches (for a review of earlier work, see Ref.~\cite{ODonnell:1981hgt}).
   Naming just a few, these include investigations in the framework of the
quark model \cite{Anisovich:1965fkk,Becchi:1965zza},
phenomenological Lagrangians \cite{Durso:1987eg,Danilkin:2017lyn},
chiral effective Lagrangians \cite{Lutz:2008km,Gomm:1984at,Hajuj:1993px,Klingl:1996by,Benayoun:1999fv,
RuizFemenia:2003hm,Terschlusen:2012xw,Chen:2013nna,Kimura:2016xnx},
QCD sum rules \cite{Zhu:1998bm,Gokalp:2001sr,Aydin:2010zz}, and lattice QCD \cite{Woloshyn:1986pk,Crisafulli:1991pn,Shultz:2015pfa,Owen:2015fra,
Alexandrou:2018jbt}.

   In this work, we perform a comprehensive study of all radiative transitions 
between the vector-meson nonet and the pseudoscalar-meson nonet in the framework of a chiral effective Lagrangian in the vector formulation, including $1/N_c$ and quark-mass corrections of first order.
   We perform fits of the available coupling constants to experimental data and 
discuss the corresponding approximations.
   In terms of five (six) coupling constants, we obtain a reasonably good 
description of the 12 experimental decay rates.
   In Sec.~II, we describe the chiral effective Lagrangian and the mixing of 
singlet and octet fields.
   Section III contains our convention of the invariant amplitude and the 
calculation of the decay rate.
   In Sec.~IV, we present the results of our fits for different levels of 
approximation.
   Finally, in Sec.~V, we conclude with a few remarks.

\section{Effective Lagrangian}
   In this section, we discuss the leading-order (LO) Lagrangian and its 
next-to-leading-order (NLO) $1/N_c$ and quark-mass corrections.
   The pseudoscalar dynamical degrees of freedom are collected in the unitary
$3\times 3$ matrix
\begin{equation}
\label{definitionU}
 U(x)=\exp\left(i\frac{\Phi(x)}{F}\right).
\end{equation}
   In Eq.~(\ref{definitionU}), $F$ denotes the pion-decay constant in the 
three-flavor chiral limit of vanishing quark masses, $m_u=m_d=m_s=0$, and is counted as $F={\cal O}(\sqrt{N_c})$ in the large-$N_c$ 
limit \cite{Witten:1979vv}.\footnote{Here, we deviate from the often-used convention of indicating the {\it three}-flavor chiral limit by a subscript~0.}
   The Hermitian $3\times 3$ matrix
\begin{align}
\Phi& =\sum_{a=1}^{8}\lambda_{a}\phi_{a} +
            \lambda_{0}\phi_{0} = \widehat{\Phi} + \widetilde{\Phi}\nonumber\\
\label{Phiphysical}
     &=     \begin{pmatrix}
            \pi ^{0} + \frac{1}{\sqrt{3}}\eta _{8} 
            + \sqrt{\frac{2}{3}}\eta _{1} & \sqrt{2}\pi ^{+} & \sqrt{2}K^{+} \\
            \sqrt{2}\pi ^{-} & -\pi ^{0} + \frac{1}{\sqrt{3}}\eta _{8} + \sqrt{\frac{2}{3}}\eta _{1} & \sqrt{2} K^{0} \\
            \sqrt{2} K^{-} & \sqrt{2}\,\overline{K}^{0} & 
            -\frac{2}{\sqrt{3}}\eta _{8} + \sqrt{\frac{2}{3}}\eta _{1}
\end{pmatrix}
\end{align}
contains the pseudoscalar octet fields $\pi, K, \eta_8$ and the pseudoscalar 
singlet field $\eta_1$, the $\lambda_a$ ($a=1,\ldots,8$) are the Gell-Mann matrices, and $\lambda_0\equiv\sqrt{2/3}\, {\mathbbm 1}$.
   In this work, we describe the vector-meson degrees of freedom within the 
so-called vector-field formalism
\cite{Ecker:1989yg,Weinberg:1995mt,Djukanovic:2010tb}.
   To that end we collect the vector fields in a Hermitian $3\times 3$ 
matrix similar to Eq.~(\ref{definitionU}),\footnote{Note that we include an additional factor 1/2.}
\begin{align}
V_{\mu}&= \left(\sum_{a=1}^{8}\frac{\lambda_{a}}{2}V_{a} 
+ \frac{\lambda_{0}}{2}V_{0} \right)_{\mu} =
    \widehat{V}_{\mu} + \widetilde{V}_{\mu} \nonumber \\
\label{Vphysical}
&=   \frac{1}{2}   \begin{pmatrix}
        \rho^{0} + \frac{1}{\sqrt{3}}\omega_{8} + \sqrt{\frac{2}{3}}\omega_{1} & \sqrt{2}\rho^{+} & \sqrt{2}K^{\ast +} \\
        \sqrt{2}\rho^{-} & -\rho^{0} + \frac{1}{\sqrt{3}}\omega_{8} + \sqrt{\frac{2}{3}}\omega_{1} & \sqrt{2} K^{\ast 0} \\
        \sqrt{2} K^{\ast -} & \sqrt{2}\,\overline{K}^{\ast 0} & 
        -\frac{2}{\sqrt{3}}\omega_{8} + \sqrt{\frac{2}{3}}\omega_{1}
        \end{pmatrix}_{\mu}.
\end{align}

   In order to construct a chirally invariant Lagrangian, we follow Gasser 
and Leutwyler by promoting the global $\mbox{U}(3)_L\times\mbox{U}(3)_R$ symmetry of QCD to a local one \cite{Gasser:1984gg}
(see, e.g., Ref~\cite{Scherer:2012zzd} for a discussion).
   In this process, we introduce external fields $s$, $p$, $l_\mu$, and $r_\mu$ 
which are Hermitian, color-neutral $3\times 3$ matrices
coupling to the corresponding quark bilinears.
   In addition, we introduce a real field $\theta$ coupling to the winding
number density.
   Introducing $u=\sqrt{U}$, the chiral vielbein $u_\mu$ and the field-strength 
tensors $f_{\pm\mu\nu}$ are defined by \cite{Ecker:1988te,Ecker:1989yg,Scherer:2012zzd}
\begin{align}
u_{\mu}&=i\left[u^\dagger(\partial_\mu-ir_\mu)u
-u(\partial_\mu-il_\mu)u^\dagger\right],\nonumber\\
f_{\pm\mu\nu}&=uf_{L\mu\nu}u^\dagger\pm u^\dagger f_{R\mu\nu}u,
\end{align}
where $l_\mu$ and $r_\mu$ denote external fields which couple to the 
corresponding currents in three-flavor QCD \cite{Gasser:1984gg}.
   In the present work, these external fields, eventually,
will contain the electromagnetic four-vector potential, and
$f_{L\mu\nu}$ and $f_{R\mu\nu}$ are the corresponding field-strength tensors,
\begin{displaymath}
f_{L\mu\nu}=\partial_\mu l_\nu-\partial_\nu l_\mu-i[l_\mu,l_\nu],\quad
f_{R\mu\nu}=\partial_\mu r_\nu-\partial_\nu r_\mu-i[r_\mu,r_\nu].
\end{displaymath}

\subsection{Lagrangian of the pseudoscalar mesons}
   We first specify the Lagrangian of the pseudoscalar sector which is relevant 
at next-to-leading order (see Ref.~\cite{Bickert:2016fgy} for more details). 
   The effective Lagrangian is organized as a {\it simultaneous} expansion in
terms of momenta $p$, quark masses $m$, and $1/N_c$.
   The three expansion variables are counted as small quantities of order
\cite{Leutwyler:1996sa}
\begin{equation}
   \label{powerexp}
   p=\mathcal{O}(\sqrt{\delta}),\ \ \ m=\mathcal{O}(\delta),\ \ \
   1/N_c=\mathcal{O}(\delta),
\end{equation}
where $\delta$ denotes a common expansion parameter.
   It is understood that dimensionful quantities such as $p$ and $m$ need to be 
small in comparison with an energy scale.
   We only specify the terms appearing in the calculation of the masses, the 
wave function renormalization constants, the decay constants, and the mixing \cite{Bickert:2016fgy}.
   The leading-order Lagrangian is given by
\cite{Leutwyler:1996sa,Kaiser:2000gs}
\begin{equation}
   \label{eq:lolagrangian}
   \mathcal{L}^{(0)}=\frac{F^2}{4}\langle D_\mu U D^\mu U^\dagger\rangle
   +\frac{F^2}{4}\langle\chi U^\dagger+U\chi^\dagger\rangle
   -\frac{1}{2}\tau\left(\sqrt{6}\frac{\eta_1}{F}+\theta\right)^2,
\end{equation}
where the symbol $\langle\ \rangle$ denotes the trace over flavor indices.
   The covariant derivatives are defined as
\begin{equation}
   \begin{split}
   \label{covariant_derivatives}
   D_\mu U&=\partial_\mu U-ir_\mu U+iU l_\mu,\\
   D_\mu U^\dagger&=\partial_\mu U^\dagger+iU^\dagger r_\mu-il_\mu U^\dagger.
   \end{split}
\end{equation}
   In Eq.~(\ref{eq:lolagrangian}), $\chi=2B_0(s+ip)$ contains the external
scalar and pseudoscalar fields \cite{Gasser:1984gg}.
   The low-energy constant (LEC) $B_0$ is related to the scalar singlet quark 
condensate $\langle\bar{q}q\rangle_0$ in the three-flavor chiral limit and is of
${\cal O}(N_c^0)$ \cite{Leutwyler:1996sa}.
   For the purposes of this work we replace $\chi\to 2B_0 {\cal M}$, where 
${\cal M}=\text{diag}(m_u,m_d,m_s)$ is the quark-mass matrix.
   Moreover, we set $\theta=0$.
   The constant $\tau=\mathcal{O}(N_c^0)$ is the topological susceptibility of
the purely gluonic theory \cite{Leutwyler:1996sa}.
   Counting the quark mass as ${\cal O}(p^2)$, the first two terms of 
$\mathcal{L}^{(0)}$ are of $\mathcal{O}(N_c p^2)$, while the third term is of $\mathcal{O}(N_c^0)$, i.e.,~all terms are of ${\cal O}(\delta^{0})$.\footnote{
	The pseudoscalar fields $\phi_0(x),\ldots, \phi_8(x)$ count as 
${\cal O}(\sqrt{N_c})$ such that in combination with
$F={\cal O}(\sqrt{N_c})$ the matrix $U$ is of ${\cal O}(N_c^0)$.}
   
   The relevant terms of the next-to-leading-order Lagrangian are given by 
\cite{Kaiser:2000gs}
\begin{align}
\label{L1}
\mathcal{L}^{(1)}&= L_5 \langle D_\mu U D^\mu U^\dagger 
(\chi U^\dagger + U \chi^\dagger)\rangle
+ L_8 \langle \chi U^\dagger\chi U^\dagger + U \chi^\dagger U \chi^\dagger \rangle  \nonumber \\
&\quad + \frac{1}{2} \Lambda_1 D_\mu\eta_1 D^\mu \eta_1 
- i \frac{F^2}{12} \Lambda_2 \left(\sqrt{6}\frac{\eta_1}{F}+\theta\right)
\langle \chi U^\dagger - U \chi^\dagger \rangle+\dots,
\end{align}
where
\begin{align}
\label{Dmueta1}
D_\mu\eta_1&=\partial_\mu\eta_1-\sqrt{\frac{2}{3}}F\langle a_\mu\rangle,\\
\label{amu}
a_\mu&=\frac{1}{2}(r_\mu-l_\mu),
\end{align}
and the ellipsis refers to the suppressed terms.
   The LECs $L_5$ and $L_8$ are of ${\cal O}(N_c)$ \cite{Gasser:1984gg} such 
that the first two terms of $\mathcal{L}^{(1)}$ count as $\mathcal{O}(N_cp^4)$.
   The LECs $\Lambda_1$ and $\Lambda_2$ represent quantities of 
${\cal O}(N_c^{-1})$ \cite{Kaiser:2000gs}.
	Therefore, all expressions of ${\cal L}^{(1)}$ are of order 
${\cal O}(\delta)$.

\subsection{Lagrangian of the vector mesons}  
    In the present case we are not interested in the interaction of vector 
mesons among each other. 
    Introducing the chiral covariant derivative of the vector-meson fields as
\begin{equation}
D_\mu V_\nu=\partial_\mu V_\nu+[\Gamma_\mu,V_\nu],
\end{equation}
where the chiral connection is given by \cite{Scherer:2012zzd}
\begin{equation}
\Gamma_\mu=\frac{1}{2}\left[u^\dagger(\partial_\mu-ir_\mu)u
+u(\partial_\mu-il_\mu)u^\dagger\right],
\end{equation}
we define the field-strength tensor as
\begin{equation}
V_{\mu\nu}=D_\mu V_\nu-D_\nu V_\mu.
\end{equation}
   The leading-order Lagrangian is then given by
\begin{equation}
{\cal L}_V=-\frac{1}{2}\langle V_{\mu\nu}V^{\mu\nu}\rangle+m_V^2\langle V_\mu V^\mu\rangle,
\end{equation}
where $m_V$ denotes the leading-order mass common to all vector-meson fields.
   We now include NLO corrections to the mass terms of ${\cal O}(N_c^{-1})$ and
${\cal O}(m)$, respectively,\footnote{For the sake of simplicity, we do not include corrections of the kinetic term. $\Delta m_S^2$ and $c_\chi$ are of order ${\cal O}(N_c^{-1})$ and ${\cal O}(N_c^0)$, respectively.}
\begin{equation}   
{\cal L}=\frac{1}{3}\Delta m_S^2\langle V_\mu\rangle\langle V^\mu\rangle
+\frac{c_\chi}{2}\langle\chi_+V_\mu V^\mu\rangle,
\end{equation}
where $\chi_+$ is defined as 
\begin{equation}
\label{chiplus}
\chi_+=u^\dagger\chi u^\dagger+u\chi^\dagger u.
\end{equation}

\subsection{Leading-order interaction Lagrangian}
    In terms of these building blocks, the leading-order Lagrangian, giving 
rise to the $VP\gamma$ interaction, is given by
\begin{equation}
\label{leading_order Lagrangian}
\mathcal{L}_{\text{LO}}=c_{1}\epsilon^{\mu \nu \rho \sigma}
\langle f_{+\mu \nu}\left\lbrace V_{\rho},u_{\sigma}\right\rbrace\rangle,
\end{equation}
where $\epsilon_{0123}=1$.
   Since $\epsilon^{\mu\nu\rho\sigma}$ and $u_\sigma$ are Lorentz pseudotensors 
of rank 4 and 1, respectively, and $f_{+\mu \nu}$ and $V_\rho$ are Lorentz tensors of rank 2 and 1, respectively, the Lagrangian of Eq.~(\ref{leading_order Lagrangian}) is even under parity.
   Moreover, the anticommutator is required in 
Eq.~(\ref{leading_order Lagrangian}) to generate positive
charge-conjugation parity (see, e.g., Ref.~\cite{Scherer:2012zzd} for more details).

   In order to describe the coupling to an external electromagnetic field, we 
insert $l_\mu=r_\mu=-eQ A_\mu$, where $e>0$ is the proton charge, $Q$ 
denotes the quark-charge matrix, and $A_\mu$ is the electromagnetic four-vector potential.
   Regarding the large-$N_c$ behavior of $Q$, we make use of the form proposed 
by B\"ar and Wiese \cite{Bar:2001qk}.
They pointed out that, when considering the electromagnetic interaction of quarks with an arbitrary number of colors, the cancelation of triangle anomalies
in the large-$N_c$ Standard Model requires the following replacement of
the ordinary quark-charge matrix,
\begin{equation}
\label{Q}
Q=\begin{pmatrix}
\frac{2}{3}&0&0\\
0&-\frac{1}{3}&0\\
0&0&-\frac{1}{3}
\end{pmatrix}
\to
\begin{pmatrix}
\frac{1}{2N_c}+\frac{1}{2}&0&0\\
0&\frac{1}{2N_c}-\frac{1}{2}&0\\
0&0&\frac{1}{2N_c}-\frac{1}{2}
\end{pmatrix}
=\begin{pmatrix}
\frac{1}{2}&0&0\\
0&-\frac{1}{2}&0\\
0&0&-\frac{1}{2}
\end{pmatrix}
+\frac{1}{2N_c}{\mathbbm 1}\equiv Q_0+Q_1.
\end{equation}
   Expanding the building blocks in the Goldstone-boson fields and keeping only 
the linear term in the expansion amounts to the replacements
\begin{equation}
\label{replacement}
f_{+\mu\nu}\to -2eQF_{\mu\nu},\quad u_\sigma\to-\frac{\partial_\sigma\Phi}{F},
\end{equation}
where $F_{\mu\nu}=\partial_\mu A_\nu-\partial_\nu A_\mu$ is the electromagnetic 
field-strength tensor.
   Thus, the LO $VP\gamma$ interaction Lagrangian, obtained from a nonlinearly 
realized chiral symmetry, reads
\begin{equation}
\label{LagrangianLOPVgamma}
{\cal L}^{VP\gamma}_{\text{LO}}=2e\frac{c_1}{F}\epsilon^{\mu\nu\rho\sigma} 
F_{\mu\nu}\langle Q\{V_\rho,\partial_\sigma\Phi\}\rangle.
\end{equation}
   The expansion of Eq.~(\ref{LagrangianLOPVgamma}) in terms of the singlet and 
octet fields is given in Appendix \ref{appendix_Lagrangians}.
   When inserting Eq.~(\ref{Q}) for the quark-charge matrix into 
Eq.~(\ref{LagrangianLOPVgamma}), we obtain the leading-order contribution proportional to $Q_0$ and a $1/N_c$ correction proportional to $Q_1$.
   When discussing our results in Sec.~IV, we will keep both scenarios in mind, 
i.e., we will compare the results obtained from using the physical quark-charge matrix $Q$ with $N_c=3$ with the expanded version truncated at order $1/N_c$ and putting $N_c=3$ at the end.

\subsection{Next-to-leading-order interaction Lagrangian}
   The NLO $1/N_c$ corrections to the Lagrangian of 
Eq.~(\ref{leading_order Lagrangian}) are obtained in terms of expressions involving two flavor traces of the same building blocks (see, e.g., Refs.~\cite{Bhaduri:1988gc,Manohar:1998xv,Bickert:2016fgy}
for an introduction to the large-$N_c$ counting),\footnote{According to Ref.~\cite{Manohar:1998xv}, the leading contribution to a correlation function of quark bilinears is of order $N_c$ and contains a single quark loop. 
	The summation over the quark flavors running in the loop amounts to taking 
a single flavor trace over the product of (flavor) $\lambda$ matrices that belong to the quark bilinears. 
	Therefore, the leading-order terms of the effective Lagrangian are also 
expected to be single-trace terms. 
	Similarly, diagrams with two quark loops have two flavor traces and are
down by one order of $1/N_c$. 
	Accordingly, double-trace terms in the effective Lagrangian are expected 
to be suppressed by one order of $1/N_c$.
	A subtlety arises because of the so-called trace relations 
\cite{Fearing:1994ga} relating linear combinations of single-trace and multiple-trace terms such that the naive counting may require a more thorough analysis (see Ref.~\cite{Gasser:1983yg}, Sec.~13).}
\begin{equation}
\label{NLO1overNcLagrangian}
\mathcal{L}_{\text{NLO,$1/N_c$}}= c_{2}\epsilon^{\mu \nu \rho \sigma}\langle V_{\rho}\rangle\langle f_{+\mu \nu}u_{\sigma}\rangle
+c_{3}\epsilon^{\mu \nu \rho \sigma}
\langle f_{+\mu \nu}V_{\rho}\rangle\langle u_{\sigma}\rangle
+c_4 \epsilon^{\mu \nu \rho \sigma}\langle f_{+\mu \nu}\rangle\langle V_{\rho} u_{\sigma}\rangle.
\end{equation}
   Performing the replacements of Eq.~(\ref{replacement}), we obtain from 
Eq.~(\ref{NLO1overNcLagrangian}) the $1/N_c$ correction to
the $VP\gamma$ interaction Lagrangian,
\begin{equation}
\label{LagrangianNLOPVgammaLNc}
{\cal L}^{VP\gamma}_{\text{NLO,$1/N_c$}}=
2\frac{e}{F}\epsilon^{\mu\nu\rho\sigma} F_{\mu\nu}
\big(c_2\langle V_\rho\rangle\langle Q\partial_\sigma\Phi\rangle
+c_3\langle QV_\rho\rangle\langle \partial_\sigma\Phi\rangle
+c_4\langle Q\rangle\langle V_\rho\partial_\sigma\Phi\rangle\big).
\end{equation}
   The first ($c_2$) term contributes to the singlet vector meson transitions,
the second ($c_3$) term to the singlet pseudoscalar transitions, and
the last ($c_4$) term vanishes for physical quark charges, because 
$\langle Q\rangle=0$ in this case.
   For the expressions in terms of the singlet and octet fields, see Appendix 
\ref{appendix_Lagrangians}.
   For $N_c=3$, the Lagrangians of Eqs.~(\ref{LagrangianLOPVgamma}) and 
(\ref{LagrangianNLOPVgammaLNc}) do not generate a singlet-to-singlet transition.
   This is a result of SU(3) symmetry \cite{Durso:1987eg}, because the 
electromagnetic current operator, consisting of octet components, cannot couple a singlet to a singlet.
   This argument no longer works for general $N_c$, because the electromagnetic 
current operator now also develops a singlet component.

   Finally, we consider quark-mass corrections in terms of the building blocks
\begin{equation}
\chi_{\pm}=u^\dagger \chi u^\dagger\pm u\chi^\dagger u.
\end{equation}
   Considering only single-trace terms, the quark-mass corrections are given by
\begin{align}
\label{NLOchiLagrangian}
\mathcal{L}_{\text{NLO,$\chi$}}
&=c_5\epsilon^{\mu\nu\rho\sigma}\langle\chi_+ f_{+\mu \nu}\lbrace V_{\rho},u_{\sigma}\rbrace\rangle
+c_6\epsilon^{\mu\nu\rho\sigma}\langle\chi_+ V_{\rho} f_{+\mu \nu} u_{\sigma} 
+\chi_{+}u_{\sigma}f_{+\mu \nu}V_{\rho}\rangle\nonumber\\
&\quad+ic_7\epsilon^{\mu\nu\rho\sigma}\langle \{f_{+\mu\nu},\partial_\rho V_\sigma\}\chi_-\rangle
+ic_8 \epsilon^{\mu\nu\rho\sigma}\langle 
f_{-\mu\nu}[\partial_\rho V_\sigma,\chi_+]\rangle.
\end{align}
   Again, making the replacements of Eq.~(\ref{replacement}), in combination 
with
\begin{equation}
\label{replacements2}
\chi_+\to 4B_0{\cal M},\quad \chi_-\to -2i \frac{B_0}{F}\{{\cal M},\Phi\},\quad
f_{-\mu\nu}\to i \frac{e}{F} F_{\mu\nu}[Q,\Phi],
\end{equation}
and performinga partial integration, we obtain from Eq.~(\ref{NLOchiLagrangian}) the first-order quark-mass correction to
the $VP\gamma$ interaction Lagrangian,
\begin{equation}
\label{LagrangianNLOPVgammachi}
{\cal L}^{VP\gamma}_{\text{NLO,$\chi$}}=4B_0\frac{e}{F}
\epsilon^{\mu\nu\rho\sigma}F_{\mu\nu}\big(
c_+\langle \{Q,V_\rho\}\{{\cal M},\partial_\sigma\Phi\}\rangle
+c_-\langle[Q,\partial_\sigma\Phi][V_\rho,{\cal M}]\rangle\big),
\end{equation}
where $c_+=c_5+c_6-c_7$ and $c_-=c_5-c_6-c_8$.
   As we will see later on, the $c_-$ term contributes only to the radiative 
transition of the $K^{\ast\pm}$.

   At this stage, we have collected the relevant Lagrangians including the 
leading $1/N_c$ and quark-mass corrections.
   Note that we consider corrections of the type $1/N_c\times\chi$ as of 
higher order.

\subsection{Field renormalization and mixing}
   Before turning to the evaluation of the transition matrix element, we need
to address two issues.
   First, the Lagrangians of the previous sections were expressed in terms of
bare fields. 
   Although we are only working at the tree level, the terms proportional to 
$L_5$ and $\Lambda_1$ contribute to the field renormalization constants.
   Second, the breaking of SU(3) symmetry due to the quark masses as well as 
the chiral anomaly generate a mixing of the singlet and octet fields.
   We neglect effects from isospin symmetry breaking, i.e., we work in the 
isospin symmetric limit $m_u=m_d=\hat m$.

   To the order we are considering, the connection between the bare pion/kaon 
fields $\phi_i$  and the renormalized pion/kaon fields $\phi_i^R$ is given by
\begin{equation}
	\label{wavefunctionrenormalization}
	\begin{split}
   \phi_i&=\sqrt{Z_\pi}\,\phi_i^R,\quad \sqrt{Z_\pi}
   =1-4 \frac{\mathring{M}_\pi^2}{F^2}L_5,\quad i=1,2,3,\\
   \phi_i&=\sqrt{Z_K}\,\phi_i^R,\quad \sqrt{Z_K}=1-4 \frac{\mathring{M}_K^2}{F^2}L_5,
   \quad i=4,5,6,7, 
   \end{split}
\end{equation}
where $\mathring{M}_{\pi}^2=2B_0\hat m$ and $\mathring{M}^2_K=(m_s+\hat m)B_0$ 
denote the lowest-order predictions for the squared pion and kaon masses,
respectively.
   For the expression of the mixing of the pseudoscalar fields, we make use of
the results of Ref.~\cite{Bickert:2016fgy}.
   Denoting the bare fields by $\eta_1$ and $\eta_8$ and the renormalized
physical fields by $\eta^R$ and $\eta'^R$, we make use of
\begin{equation}
\label{mixing_pseudoscalars}
\begin{split}
\eta_8&=\left[\left(1-\frac{1}{2}\delta_8\right)\cos(\theta_P)
+\frac{1}{2}\delta_{81}\sin(\theta_P)\right]\eta^R
+\left[\left(1-\frac{1}{2}\delta_8\right)\sin(\theta_P)
-\frac{1}{2}\delta_{81}\cos(\theta_P)\right]\eta'^R,\\
\eta_1&=\left[-\frac{1}{2}\delta_{81}\cos(\theta_P)
-\left(1-\frac{1}{2}\delta_1\sin(\theta_P)\right)\right]\eta^R
+\left[-\frac{1}{2}\delta_{81}\sin(\theta_P)
+\left(1-\frac{1}{2}\delta_1\right)\cos(\theta_P)\right]\eta'^R,
\end{split}
\end{equation}
where
\begin{align*}
\delta_8&=\frac{8(4\mathring{M}_K^2-\mathring{M}_\pi^2)}{3F^2}L_5,\\
\delta_1&=\frac{8(2\mathring{M}_K^2+\mathring{M}_\pi^2)}{3F^2}L_5+\Lambda_1,\\
\delta_{81}&=-\frac{16\sqrt{2}(\mathring{M}_K^2-\mathring{M}_\pi^2)}{3F^2}L_5.
\end{align*}
	Using the numerical values for the masses and low-energy constants from the 
next subsection, we obtain for the pseudoscalar mixing angle the values 
$\theta^{[0]}_P=-19.7\degree$ and $\theta^{[1]}_P=-12.4\degree$
at leading order and next-to-leading order, respectively.
	 These values are representative and cover the range for $\theta_P$ between 
$-10\degree$ and $-20\degree$ reported in Ref.~\cite{Zyla:2020zbs}.
   
   In the case of the vector mesons, we only consider $\phi$-$\omega$ mixing in
the form  
\begin{equation}
\label{mixingV}
\begin{pmatrix}
\phi\\
\omega
\end{pmatrix}
=\begin{pmatrix}
\cos(\theta_V)&-\sin(\theta_V)\\
\sin(\theta_V)&\cos(\theta_V)
\end{pmatrix}
\begin{pmatrix}
\omega_8\\
\omega_1
\end{pmatrix}
\equiv R_V
\begin{pmatrix}
\omega_8\\
\omega_1
\end{pmatrix}.
\end{equation}
   The diagonal mass matrix of the physical fields is related to the symmetric 
mass matrix in the octet-singlet basis, including the NLO corrections of ${\cal O}(N_c^{-1})$ 
and ${\cal O}(m)$, via 
\begin{equation}
{\cal M}^2_{V,\rm phys}=
\begin{pmatrix}
m^2_\phi&0\\
0&m^2_\omega
\end{pmatrix}
=R_V\begin{pmatrix}m^2_8&m^2_{81}\\m^2_{81}&m_1^2\end{pmatrix}R^T_V,
\end{equation}
where, to the order we are working at,
\begin{align*}
m_8^2&=m_V^2+\frac{c_\chi}{3}(4\mathring{M}_K^2-\mathring{M}_\pi^2),\\
m_1^2&=m_V^2+\Delta m_S^2+\frac{c_\chi}{3}(\mathring{M}_\pi^2+2\mathring{M}_K^2),\\
m_{18}^2&=-\frac{2\sqrt{2}c_\chi}{3}(\mathring{M}_K^2-\mathring{M}_\pi^2).
\end{align*}
   The mixing angle is obtained from the relation
\begin{equation}
\label{tanthetaV}
\tan(\theta_V)=\sqrt{\frac{m_\phi^2-m_8^2}{m_8^2-m_\omega^2}},
\end{equation}
where $m_8^2$ satisfies, to the order we are working at,
\begin{displaymath}
   m_8^2=\frac{1}{3}(4m^2_{K^\ast}-m_\rho^2),
\end{displaymath}
resulting in 
\begin{displaymath}
\tan(\theta_V)=\sqrt{\frac{3m_\phi^2+m_\rho^2-4m_{K^\ast}^2}{4m_{K^\ast}^2
		-m_\rho^2-3m_\omega^2}}.
\end{displaymath}
   For the mixing angle we obtain $\theta_V=39.8\degree$, which turns out
to be close to the ideal mixing $\theta_V=35.3\degree$, corresponding to $\phi=-s\bar s$ and $\omega=(u\bar u+d\bar d)/\sqrt{2}$ in the quark model,
\begin{equation}
   \label{idealmixing}
   \begin{split}
   \phi_\text{ideal}&=\sqrt{\frac{2}{3}}\,\omega_8-\frac{1}{\sqrt{3}}\,\omega_1,\\
   \omega_\text{ideal}&=\frac{1}{\sqrt{3}}\,\omega_8+\sqrt{\frac{2}{3}}\,\omega_1.
   \end{split}
\end{equation}

\subsection{Numerical values for masses and parameters}
   For the empirical masses of the pseudoscalar mesons and the vector mesons we 
make use of the values given in Table \ref{table_masses} \cite{Zyla:2020zbs}.
	For the decay constants we take $F_{\pi}=92.2\,\text{MeV}$ and 
$F_K=110.\,\text{MeV}$ \cite{Zyla:2020zbs}.\footnote{Here and in the following, 
an integer followed by a point denotes a rounded number rather than an exact integer.}
\renewcommand{\arraystretch}{1.3}
\begin{table}[p]
	\caption{Masses of the pseudoscalar mesons and the vector mesons in MeV.}
	\label{table_masses}
	\begin{center}
		\begin{tabular}{cccccc}
			\hline
			\hline
			$M_{\pi^\pm}$\quad&\quad$M_{\pi^0}$\quad&\quad$M_{K^\pm}$\quad&
			\quad$M_{K^0/\overline{K}^0}$\quad&\quad$M_\eta$\quad&
			\quad$M_{\eta'}$\\
			139.6&\quad 135.0&\quad 493.7&\quad 497.6&\quad 547.9&\quad 957.8\\
			\hline
			$m_{\rho^\pm}$\quad&\quad$m_{\rho^0}$\quad&\quad$m_{K^{\ast\pm}}$
			\quad&\quad$m_{K^{\ast 0}/\overline{K}^{\ast 0}}$\quad&\quad$m_\omega$\quad&
			\quad$m_{\phi}$\\
			775.1&\quad 775.3&\quad891.8&\quad895.6&\quad782.7&\quad1019.5\\
			\hline
			\hline
		\end{tabular}
	\end{center}
\end{table}
\renewcommand{\arraystretch}{1}
   The predictions for the squared pion and kaon masses are obtained from the
one-loop expressions of chiral perturbation theory 
\cite{Gasser:1984gg,Scherer:2002tk} by dropping the loop contributions and the tree-level contributions proportional to $L_6$ and $L_4$,
\begin{align*}
M_\pi^2&=2B\hat{m}\left[1+\frac{16B\hat m}{F^2}(2L_8-L_5)\right],\\
M_K^2&=B(m_s+\hat m)\left[1+\frac{8B(m_s+\hat m)}{F^2}(2L_8-L_5)\right].
\end{align*} 
   In terms of the quark mass ratio $r$ \cite{Zyla:2020zbs}, 
\begin{equation}
r = \frac{m_{s}}{\hat{m}} = 27.37,
\end{equation}
we obtain for the lowest-order squared pion and kaon masses
\begin{align}
\begin{split}
\mathring{M}_{\pi}^{2} =& \dfrac{r+1}{r-1}\overline{M}_{\pi}^{2} 
+ 4\dfrac{\overline{M}_{K}^{2}}{1-r^{2}},
\\
\mathring{M}_{K}^{2} =& 
\dfrac{\left( 1+r \right)^{2}\overline{M}_{\pi}^{2} - 4 \overline{M}_{K}^{2}}
{2\left( r-1 \right)},
\end{split}
\end{align}
where
\begin{displaymath}
\overline{M}_{\pi} = \dfrac{M_{\pi ^{0}}+M_{\pi ^{\pm}}}{2} \quad \mathrm{and}
\quad
\overline{M}_{K} =\dfrac{M_{K ^{0}}+M_{K ^{\pm}}}{2}.
\end{displaymath}
	Using, in addition, the expressions for the pion and kaon decay constants 
$F_\pi$ and $F_K$,
\begin{equation}
\label{pionkaondecayconstants}
\begin{split}
F_\pi&=F\left(1+4 \frac{\mathring{M}_\pi^2}{F^2}L_5\right),\\
F_K&=F\left(1+4\frac{\mathring{M}_K^2}{F^2}L_5\right),
\end{split}
\end{equation}
we can write
\begin{align}
\begin{split}
F =& \dfrac{\mathring{M}_{K}^{2}F_{\pi} - \mathring{M}_{\pi}^{2}F_{K}}
{\mathring{M}_{K}^{2}-\mathring{M}_{\pi}^{2}},
\\
L_{5} =& \dfrac{F\left( F_{\pi} - F \right)}{4\mathring{M}_{\pi}^{2}},
\\
L_{8} =& \dfrac{F^{2}}{4\left( 1 - r^{2} \right)\mathring{M}_{\pi}^{4}}\left( \dfrac{1+r}{2}\overline{M}_{\pi}^{2} - \overline{M}_{K}^{2} \right) 
+ \dfrac{L_{5}}{2}.
\end{split}
\end{align}
   The corresponding values for $\mathring{M}_\pi$, $\mathring{M}_K$,
$F$, $L_5$, and $L_8$ are given in Table \ref{Tab:WerteGroessenImModell}.
\renewcommand{\arraystretch}{1.3}
\begin{table}[h]
	\begin{center}
		\caption{Numerical values of lowest-order pion and kaon masses and LECs .}
		\label{Tab:WerteGroessenImModell}
		\begin{tabular}{ccccc}
			\hline
			\hline
			$\mathring{M}_{\pi}$ & $\mathring{M}_{K}$ & $F$ &
			$L_{5}$ & $L_{8}$ 
			\\
			$137.7 ~\mathrm{MeV}$ & $518.7 ~\mathrm{MeV}$ & $90.9 ~\mathrm{MeV}$
			&		
			$1.62\cdot 10^{-3}$ & $0.642\cdot 10^{-3}$ 
			\\
			\hline
			\hline
		\end{tabular}
	\end{center}
\end{table}
\renewcommand{\arraystretch}{1.}

\section{Invariant matrix element and decay rate}
   The invariant amplitude of the decay $V(p,\epsilon_V)\to 
P(k)+\gamma(q,\epsilon)$ may be parametrized as\footnote{We follow the convention of Ref.\ \cite{Bjorken:1965sts} such that the invariant amplitude is obtained from $i{\cal L}_{\rm int}$.}
\begin{equation}
\label{Mpar}
{\cal M}=-2i e {\cal A} \epsilon^{\mu\nu\rho\sigma} q_\mu \epsilon_\nu^\ast(q)\epsilon_{V\rho}(p)k_\sigma,
\end{equation}
where four-momentum conservation $p=k+q$ is implied, $\epsilon$ and 
$\epsilon_V$ denote the polarization vectors of the photon and the vector meson, respectively, and the amplitude $\cal A$ is determined from the Lagrangians of Eqs.~(\ref{LagrangianLOPVgamma}),
(\ref{LagrangianNLOPVgammaLNc}), and (\ref{LagrangianNLOPVgammachi}).
   The invariant amplitude for the decay $P(k)\to V(p,\epsilon_V)+\gamma(q,\epsilon)$ is obtained 
from Eq.~(\ref{Mpar}) by substituting $\epsilon_V\to\epsilon_V^\ast$ and 
$k\to -k$.

    In the rest frame of the initial-state particle, the differential decay 
rate for the decay $A(p_A)\to B(p_B)+\gamma(q)$ is given by \cite{Bjorken:1965sts}
\begin{equation}
d\Gamma = \dfrac{1}{2m_{A}}\overline{|\mathcal{M}|^{2}}
\dfrac{d^3p_B}{2E_B(2\pi)^3}\dfrac{d^3q}{2E_{\gamma}(2\pi)^3}
(2\pi)^{4}\delta^4(p_A-p_B-q),
\end{equation}
where $E_{B}$ and $E_{\gamma}$ denote the energies of the decay product $B$ 
and the real photon, respectively.
   When averaging over the initial polarizations and summing over the final 
polarizations, we make use of the ``completeness relations'' \cite{Ryder:1985wq} for the polarization vectors of the photon and the vector meson, respectively,\footnote{As
usual it is assumed that the photon polarization vector is contracted with the matrix element of the conserved electromagnetic current.}
\begin{align*}
\sum_{\lambda = \pm 1}\epsilon_\nu^{\ast}(q,\lambda)\epsilon_{\nu '}(q,\lambda)
&=-g_{\nu \nu '},\\
\sum_{\lambda=-1}^{+1} \epsilon_{V\rho}(p,\lambda)
\epsilon_{V\rho '}^{\ast}(p,\lambda)& =
\left(-g_{\rho\rho '}+\dfrac{p_{\rho}p_{\rho '}}{m_{V}^{2}}\right),
\end{align*}
where $m_{V}$ is the mass of the vector meson.
   Using \cite{Itzykson:1980rh}
\begin{displaymath}
\epsilon^{\mu\nu\rho\sigma}{\epsilon^{\mu'}}_{\nu}\,^{\rho'\sigma'}
=-\text{det}(g^{\alpha\alpha'}),\quad\alpha=\mu,\rho,\sigma,\quad
\alpha'=\mu',\rho',\sigma',
\end{displaymath}
in combination with the on-shell conditions $p^2_A=m_A^2$, $p_B^2=m_B^2$, 
and $q^2=0$, we obtain
\begin{displaymath}
\overline{|\mathcal{M}|^{2}}=c_A 2e^2|{\cal A}|^2(m_A^2-m_B^2)^2,
\end{displaymath}
where $c_A=1/3$ for a vector meson in the initial state and $c_A=1$ for a 
pseudoscalar meson in the initial state.
   Using \cite{Ryder:1985wq}
\begin{displaymath}
\frac{d^3 q}{2E_\gamma}=d^4 q \delta(q^2)\Theta(q_0),
\end{displaymath}
we obtain for the decay rate
\begin{align}
\label{resultGamma}
\Gamma_{A\to B\gamma}&=
    \dfrac{1}{2m_{A}}
        \int \dfrac{d^3p_B}{2E_B(2\pi )^3}
            \dfrac{d^3q}{2E_{\gamma}(2\pi )^{3}}
            (2\pi )^{4}\delta^4(p_A-p_B-q)
            \overline{|\mathcal{M}|^{2}}\nonumber\\
    &= \dfrac{1}{16\pi ^{2}m_{A}}
        \int \dfrac{d^3 p_B}{E_B}
        \int d^4 q \delta(q^2)\Theta(q_0)
        \delta^{4}(p_A-p_B-q) \overline{|\mathcal{M}|^{2}}\nonumber \\
    &=c_A \frac{e^2|{\cal A}|^2}{8\pi}\left(\frac{m_A^2-m_B^2}{m_A}\right)^3.
\end{align}

\section{Results and discussion}
   Starting from the expression for the decay rate, Eq.~(\ref{resultGamma}), we
determine the low-energy coupling constants of the interaction Lagrangians by
fitting the corresponding expressions to the available experimental data.
   For the masses of the pseudoscalar mesons and the vector mesons we make use 
of the values given in Table \ref{table_masses}.
   The experimental partial widths were calculated with the aid of the PDG 
values of the total widths in combination with the corresponding branching
ratios \cite{Zyla:2020zbs} (see second column of Table \ref{table_fit_1}).

\subsection{Leading order}
\label{subsection_leading_order}
   In the following, we investigate different levels of approximation and 
compare the different scenarios.
   To that end, we start with the results corresponding to the leading-order 
Lagrangian of Eq.~(\ref{LagrangianLOPVgamma}) in combination with the pseudoscalar mixing angle obtained at leading order,
$\theta^{[0]}_P=-19.7\degree$,
and the vector mixing angle corresponding to ideal mixing, i.e., $\cos(\theta_V)=\sqrt{2/3}$ and $\sin(\theta_V)=1/\sqrt{3}$.
   When fitting the data, we made use of the {\it Mathematica} package 
\texttt{NonLinearModelFit} \cite{Wolfram:2016}.
   In order to facilitate identifying which decays are well-described and which 
are not, we introduce both a relative deviation and a deviation normalized with respect to the uncertainties as
\begin{equation}
\label{defd1d2}
\delta_1=\frac{\Gamma_\text{mod}-\Gamma_\text{exp}}{\Gamma_\text{exp}},\quad
\delta_2=\frac{\Gamma_\text{mod}-\Gamma_\text{exp}}{\sqrt{\sigma^2_\text{mod}
		+\sigma^2_\text{exp}}}.
\end{equation}
   Here, $\sigma_\text{exp}$ and $\sigma_\text{mod}$ denote the experimental
uncertainty and the estimated model uncertainty, respectively.
   As a rule of thumb, values for $|\delta_2|$ larger than one indicate tension 
between the model and the experimental results.
   The result of the fit to the data is shown in Table \ref{table_fit_1} with 
$|c_1|=(3.82\pm 0.25)\times 10^{-2}$.
   Note that because of the Okubo-Zweig-Iizuka (OZI) rule
\cite{Okubo:1963fa,Zweig:1964,Iizuka:1966fk}, at leading order, the decay rate 
for $\phi\to\pi^0\gamma$ vanishes as $N_c\to\infty$, independently from the value of the coupling constant $c_1$.
   Therefore, we have excluded this decay from the fit.
   Neglecting $\eta$-$\eta'$ mixing, i.e., taking $\theta_P=0\degree$, the 
leading-order Lagrangian results generate the same ratios of the magnitudes 
of the decay amplitudes as the quark model with SU(6) symmetry \cite{Anisovich:1965fkk}.

   In general, the numbers of the tables were rounded at the end of the
calculation.
   Since the decay rate is a function of $|{\cal A}|^2$, it is not possible to 
extract the sign of $c_1$.
   For the sake of simplicity, we assume $c_1>0$ such that the signs of the 
remaining coupling constants, to be determined below, will be given with 
respect to a positive $c_1$.
   Except for the decays $\omega\to\eta\gamma$ and $\phi\to\eta'\gamma$, the
theoretical partial decay widths are smaller than the experimental ones.
   Furthermore, we note that only for the decay $\omega\to\eta\gamma$ we find 
a deviation $|\delta_2|$ which is smaller than one.
   Using the experimental uncertainties, we obtain for the reduced chi-squared,
\begin{displaymath}
\chi^2_\text{red}=\frac{1}{\nu}\sum_{i=1}^{11} \frac{\left(\Gamma^\text{exp}_i-\Gamma^\text{LO}_i\right)^2}{\sigma_i^2}=94.,
\end{displaymath}
where, omitting $\phi\to\pi^0\gamma$, the number of degrees of freedom is 
$\nu=11-1=10$ at leading order.
   We conclude that a description in terms of a single coupling constant $c_1$ 
does not provide a good description of the twelve decays.
\renewcommand{\arraystretch}{1.3}
\begin{table}[h]
\caption{Comparison of the decay rates at LO with experimental values 
\cite{Zyla:2020zbs}.}
\label{table_fit_1}
\begin{tabular}{lcccc}
\hline
\hline
Decay &\quad $\Gamma_\text{exp}$ (keV) &\quad $\Gamma_\text{LO}$ (keV) 
&\quad deviation $\delta_1$&\quad deviation $\delta_2$\\
\hline
$\rho^{0}\rightarrow \pi^{0}\gamma$ &\quad $70.\pm 12.$ &\quad $40.7 \pm 5.4$ 
&\quad $-0.42$ &\quad $-2.3$\\
$\rho^{\pm}\rightarrow \pi^{\pm}\gamma$ &\quad $67.1 \pm 7.5$ 
&\quad $40.4 \pm 5.3$ &\quad $-0.40$ &\quad $-2.9$\\
$\rho^{0}\rightarrow \eta \gamma$ &\quad $44.7\pm 3.1$ &\quad $33.7 \pm 4.5$ &\quad $-0.25$ &\quad $-2.0$\\
$\omega\rightarrow \pi^{0}\gamma$ &\quad $723.\pm 25.$ &\quad $364. \pm 48.$ 
&\quad $-0.50$ &\quad $-6.6$\\
$\omega \rightarrow \eta \gamma$ &\quad $3.91 \pm 0.35$ &\quad $4.07 \pm 0.54$ &\quad $0.041$ &\quad $0.25$\\
$\phi \rightarrow \pi ^{0}\gamma$ &\quad $5.61 \pm 0.26$ &\quad - &\quad - 
&\quad -\\
$\phi \rightarrow \eta \gamma$ &\quad $55.4 \pm 1.1$ &\quad $48.1 \pm 6.4$ &\quad $-0.13$ &\quad $-1.1$\\
$\phi \rightarrow \eta ' \gamma$ &\quad $0.2643 \pm 0.0090$ 
&\quad $0.439 \pm 0.058$ &\quad $0.66$ &\quad $3.0$ \\
$K^{\ast 0}\rightarrow K^{0}\gamma$ &\quad $116. \pm 10.$ &\quad $91. \pm 12.$ &\quad $-0.22$ &\quad $-1.6$\\
$K^{\ast \pm}\rightarrow K^{\pm}\gamma$ &\quad $50.4 \pm 4.7$ 
&\quad $22.6 \pm 3.0$ &\quad $-0.55$ &\quad $-5.0$\\
$\eta ' \rightarrow \rho ^{0}\gamma$ &\quad $55.5 \pm 1.9$ 
&\quad $30.7 \pm 4.1$&\quad $-0.45$ &\quad $-5.6$\\
$\eta ' \rightarrow \omega \gamma$ &\quad $4.74 \pm 0.20$ 
&\quad $3.06 \pm 0.40$ &\quad $-0.35$ &\quad $-3.7$\\
\hline
\hline
\end{tabular}
\end{table}
\renewcommand{\arraystretch}{1}

\subsection{$1/N_c$ corrections}

   In the next scenario, we consider the $1/N_c$ corrections, but still stick 
to the SU(3) symmetry of the interaction terms.
   For the $\phi$-$\omega$ mixing we still take ideal mixing.
   Using $\overline{M}^2=M_K^2=M_\pi^2$ in the SU(3)-symmetric case, we find 
from Eq.~(\ref{LagrangianNLOPVgammaexp2app}) of
Appendix~\ref{appendix_Lagrangians} that the quark-mass corrections simply 
result in a shift of the coupling constant $c_1$ of the leading-order Lagrangian, i.e., $c_1\to \tilde{c}_1=c_1 +2\overline{M}^2c_+$.
   On the other hand, the $1/N_c$ corrections  (see Table \ref{table_Tifull} of 
Appendix~\ref{appendix_Lagrangians}) affect both the $\rho^0\eta_1$ 
and $\omega_8\eta_1$ transitions in terms of the replacement $\tilde{c}_1\to\tilde{c}_1+\frac{3}{2}c_3$ and, similarly, both the $\pi^0\omega_1$ and $\eta_8\omega_1$ transitions in terms of the replacement $\tilde{c}_1\to\tilde{c}_1+\frac{3}{2}c_2$.
	The results of the fit for the SU(3)-symmetric case are shown in Table 
\ref{table_fit_SU(3)}.
   	The reduced chi-squared is now 45.~(for twelve decays and 9 degrees of 
freedom) in comparison with 94.~of the LO fit.
   The effective coupling constant $\tilde{c}_1$ comes out as 
$\tilde{c}_1=(3.36\pm 0.20)\times 10^{-2}$.
   Therefore, the decay rates for $\rho\to\pi\gamma$ and $K^\ast\to K\gamma$, 
which are not affected by $c_2$ and $c_3$, are reduced by the factor 
$(\tilde{c}_1/c_1)^2=0.77$.
   For the other decays, the situation is more complex.
   Even though the transitions $\omega_8\to\eta_8\gamma$, 
$\omega_1\to\eta_1\gamma$, $\rho^0\to\eta_8\gamma$,
and $\omega_8\to\pi^0\gamma$ are still described in terms of $\tilde{c}_1$,
because of the mixing of Eqs.~(\ref{mixing_pseudoscalars}) and (\ref{mixingV}), 
all of the remaining physical decays beyond $\rho\to\pi\gamma$ and $K^\ast\to K\gamma$ contain $\tilde{c}_1$ as well as
$c_2=(0.67\pm 0.10)\times 10^{-2}$ and $c_3=(-0.39\pm 0.25)\times 10^{-2}$.

\renewcommand{\arraystretch}{1.3}
\begin{table}[h]
\caption{Decay rates including $1/N_c$ corrections in the SU(3)-symmetric case. 
For the experimental values, see Table \ref{table_fit_1}.}
\label{table_fit_SU(3)}
\begin{tabular}{lccc}
\hline
\hline
Decay &\quad $\Gamma_\text{LO+1/$N_c$}$ (keV) &\quad deviation $\delta_1$
&\quad deviation $\delta_2$\\
\hline
$\rho^{0}\rightarrow \pi^{0}\gamma$ &\quad $31.6 \pm 3.7$ &\quad $-0.55$ 
&\quad $-3.1$\\
$\rho^{\pm}\rightarrow \pi^{\pm}\gamma$ &\quad $31.3 \pm 3.7$ &\quad $-0.53$ &\quad $-4.3$\\
$\rho^{0}\rightarrow \eta \gamma$ &\quad $23.2 \pm 3.2$ &\quad $-0.48$ 
&\quad $-4.8$\\
$\omega\rightarrow \pi^{0}\gamma$ &\quad $400. \pm 35.$ &\quad $-0.45$ &\quad $-7.6$\\
$\omega \rightarrow \eta \gamma$ &\quad $5.64 \pm 0.63$ &\quad $0.44$ 
&\quad $2.4$\\
$\phi \rightarrow \pi ^{0}\gamma$ &\quad $5.4 \pm 1.7$ &\quad $-0.039$ 
&\quad $-0.13$\\
$\phi \rightarrow \eta \gamma$ &\quad $59.1 \pm 6.1$ &\quad $0.066$ &\quad $0.59$\\
$\phi \rightarrow \eta ' \gamma$ &\quad $0.298 \pm 0.056$ &\quad $0.13$ 
&\quad $0.60$ \\
$K^{\ast 0}\rightarrow K^{0}\gamma$ &\quad $70.5 \pm 8.4$ &\quad $-0.39$ 
&\quad $-3.5$\\
$K^{\ast \pm}\rightarrow K^{\pm}\gamma$ &\quad $17.6 \pm 2.1$ &\quad $-0.65$ &\quad $-6.4$\\
$\eta ' \rightarrow \rho ^{0}\gamma$ &\quad $46.5 \pm 4.1$&\quad $-0.16$ 
&\quad $-2.0$\\
$\eta ' \rightarrow \omega \gamma$ &\quad $5.44 \pm 0.67$ &\quad $0.15$ 
&\quad $1.0
$\\
\hline
\hline
\end{tabular}
\end{table}
\renewcommand{\arraystretch}{1}

\subsection{$1/N_c$ and quark-mass corrections}
\label{subsection_LNc_quark_mass}

	SU(3) symmetry implies that the amplitudes ${\cal A}$ of the decays 
$\rho\to\pi\gamma$, $K^{\ast\pm}\to K^\pm\gamma$, and $K^{\ast 0}\to K^0\gamma$
satisfy the relations $|{\cal A}_{\rho\to\pi\gamma}|=|{\cal A}_{K^{\ast\pm}\to K^\pm\gamma}|$ and $|{\cal A}_{K^{\ast 0}\to K^0\gamma}|=2|{\cal A}_{K^{\ast\pm}\to K^\pm\gamma}|$ \cite{ODonnell:1981hgt,Anisovich:1965fkk} (see also Table \ref{table_Tifull}).
	Using Eq.~(\ref{resultGamma}) together with the physical masses of Table 
\ref{table_masses} and the experimental decay rates of Table \ref{table_fit_1},
one obtains 
$|{\cal A}_{\rho^\pm\to\pi\gamma}|/|{\cal A}_{K^{\ast\pm}\to K^\pm\gamma}|=0.909$ and $|{\cal A}_{K^{\ast 0}\to K^0\gamma}|/|{\cal A}_{K^{\ast\pm}\to K^\pm\gamma}|=1.59$, amounting, at the amplitude level,
to an SU(3)-symmetry breaking of about 9\% and 20\%, respectively.
	This is of the same order of magnitude as the relative difference between 
the decay constants $F_\pi$ and $F_K$, $(F_K-F_\pi)/F_K=16\%$.
	The experimental decay rates for $\omega\to\pi^0\gamma$ and 
$\rho^0\to\pi^0\gamma$ result in $|{\cal A}_{\omega\to\pi^0\gamma}|/
|{\cal A}_{\rho^0\to\pi^0\gamma}|=2.88$, very close to 3, the leading-order 
large-$N_c$ prediction.

	In the next step, we include the SU(3)-symmetry-breaking terms.
	With regard to the vector mesons, we now have to consider the 
$\phi$-$\omega$ mixing at next-to-leading order with a mixing angle of $\theta_V=39.8\degree$.\footnote{Since we did not take any corrections to the 
kinetic term into account, the wave function renormalization constants are still 1 for the vector mesons.} 
	For the decays involving pions and kaons, we need to take the 
wave function renormalization constants of
Eqs.~(\ref{wavefunctionrenormalization}) into account.
	In terms of the pion and kaon decay constants of 
Eqs.~(\ref{pionkaondecayconstants}), this amounts to replacing in the 
leading-order Lagrangian of Eq.~(\ref{LagrangianLOPVgamma}) the decay constant $F$ by the physical $F_\pi$ and $F_K$ in the corresponding cases.
	With reference to the Lagrangians of Eqs.~(\ref{LagrangianNLOPVgammaLNc})
and (\ref{LagrangianNLOPVgammachi}) such replacement is of higher order.
	For the decays that involve an $\eta$ or $\eta'$, the situation is more 
complicated because of the mixing.
	Here, we make use of Eqs.~(\ref{mixing_pseudoscalars}) in combination with 
the NLO mixing angle $\theta_P^{[1]}=-12.4\degree$.
	Equations~(\ref{mixing_pseudoscalars}) introduce one additional, so far 
unspecified LEC of order $1/N_c$, namely, $\Lambda_1$, originating from the 
NLO kinetic Lagrangian of Eq.~(\ref{L1}). 
	We performed three fits with $\Lambda_1=-1/3,0,1/3$, yielding the results 
shown in Table \ref{Tab:VariationConstantLambda1}. 

\begin{table}[h]
	\renewcommand{\arraystretch}{1.3}
	\begin{center}
		\caption{Decay rates including $1/N_c$ and quark-mass corrections 
			for $\Lambda_1=-1/3,0,1/3$. For the mixing angles we made use of 
			the NLO values $\theta_V=39.8\degree$ 
			and $\theta_P^{[1]}=-12.4\degree$.}
		\label{Tab:VariationConstantLambda1}
		\begin{tabular}{lcccc}
			\hline
			\hline
			Decay &\quad $\Gamma _{\mathrm{exp}}$ (keV) &\quad $\Gamma$ (keV), 
			$\Lambda _{1} = -\frac{1}{3}$ & \quad $\Gamma$ (keV), $\Lambda _{1} = 0$ &\quad $\Gamma$ (keV), $\Lambda _{1} = \frac{1}{3}$
			\\
			\hline
			$\rho ^{0}\rightarrow \pi ^{0}\gamma$ & $70.\pm 12.$ 
			& $32. \pm 12.$ & $52.4 \pm 9.8$ & $71.5 \pm 6.4$
			\\ 
			$\rho ^{\pm}\rightarrow \pi ^{\pm}\gamma$ & $67.1 \pm 7.5$ 
			& $32. \pm 12.$ & $52.0 \pm 9.7$ & $71.0 \pm 6.4$
			\\ 
			$\rho ^{0}\rightarrow \eta \gamma$ & $44.7\pm 3.1$ 
			& $21.4 \pm 8.0$ & $37.4 \pm 6.3$ & $54.2 \pm 4.1$
			\\ 
			$\omega \rightarrow \pi ^{0}\gamma$ & $723.\pm 25.$ 
			& $299. \pm 87.$ & $459. \pm 69.$ & $610. \pm 46.$
			\\ 
			$\omega \rightarrow \eta \gamma$ & $3.91\pm 0.35$ 
			& $1.95 \pm 0.63$ & $3.32 \pm 0.50$ & $4.64 \pm 0.33$
			\\ 
			$\phi \rightarrow \pi ^{0}\gamma$ & $5.61\pm 0.26$ 
			& $4.5 \pm 2.3$ & $5.0 \pm 1.4$ & $5.65 \pm 0.77$
			\\
			$\phi \rightarrow \eta \gamma$ & $55.4\pm 1.1$ 
			& $52.0 \pm 9.7$ & $53.9 \pm 6.0$ & $56.1 \pm 3.3$
			\\
			$\phi \rightarrow \eta ' \gamma$ & $0.2643 \pm 0.0090$ 
			& $0.26 \pm 0.82$ & $0.258 \pm 0.050$ & $0.254 \pm 0.027$
			\\ 
			$K^{\ast 0}\rightarrow K^{0}\gamma$ & $116.\pm 10.$ 
			& $86. \pm 16.$ & $111. \pm 11.$ & $135.7 \pm 6.9$
			\\ 
			$K^{\ast \pm}\rightarrow K^{\pm}\gamma$ & $50.4\pm 4.7$ 
			& $50. \pm 43.$ & $50. \pm 26.$ & $50. \pm 14.$
			\\ 
			$\eta ' \rightarrow \rho ^{0}\gamma$ & $55.5\pm 1.9$ 
			& $70.1 \pm 9.1$ & $62.0 \pm 4.7$ & $48.1 \pm 2.0$ 
			\\ 
			$\eta ' \rightarrow \omega \gamma$ & $4.74\pm 0.20$ 
			& $6.6 \pm 1.3$ & $6.48 \pm 0.72$ & $5.40 \pm 0.30$
			\\
			\hline
			$\chi ^{2}_{\mathrm{red}}$ & - & 84. & 32. & 9.4
			\\
			\hline
			\hline
		\end{tabular}
	\end{center}
	\renewcommand{\arraystretch}{1}
\end{table}

   	Since the results turned out to be highly dependent on the value of the 
$\Lambda_1$ parameter, we performed fits that included $\Lambda_1$ as a 
free parameter in addition to the $c_i$ parameters.
	In this context, we also consider two different scenarios: in the first 
case (denoted by I) we calculate the amplitude up to and including NLO and fit 
its square, whereas in the second case (II) we fit the squared amplitude only up to and including NLO. 
	In other words, in the second case we do not keep terms of the order 
$\mathrm{NNLO} = \mathrm{NLO} \times \mathrm{NLO}$ in the decay rate.  
    Omitting for notational convenience the summation or averaging over the 
spins, we thus consider\footnote{Our previous results thus correspond to the 
first case.}
\begin{enumerate}
	\item[I:] $\left\vert \mathcal{M}^\text{I} \right\vert ^{2} = \left\vert \mathcal{M}_\text{LO} \right\vert ^{2} + 2 \mathrm{Re}\left( \mathcal{M}_\text{LO}\mathcal{M}^{\ast}_\text{NLO} \right) + \left\vert \mathcal{M}_\text{NLO} \right\vert ^{2}$,
	\item[II:] $\left\vert \mathcal{M}^\text{II} \right\vert ^{2} = \left\vert \mathcal{M}_\text{LO} \right\vert ^{2} + 2 \mathrm{Re}\left( \mathcal{M}_\text{LO}\mathcal{M}^{\ast}_\text{NLO} \right)$.
\end{enumerate}
	The results for the two fits are shown in Table \ref{Tab:VariationMatrElem}.
	Judging from the value of the reduced chi-squared, 
$\chi^2_{\text{red}}=6.1$, we conclude that the second method provides the 
best description of the data.
	The corresponding set of parameters is given by
\begin{equation}
	\begin{split}
	c_{1}& = 0.0522 \pm 0.0020,\quad  c_{2} = -0.00100 \pm 0.00021
	,\quad
	c_{3} = 0.00272 \pm 0.00072,\\
	c_{+}& = \left(2.80 \pm 0.78 \right)\cdot 10^{-9}\,\mathrm{MeV}^{-2},
	\quad c_{-} = \left( -6,1 \pm 43.9 \right)\cdot 10^{-9} \,\mathrm{MeV}^{-2},
	\quad \Lambda _{1} = 0.290 \pm 0.050.
	\end{split}
\end{equation}
	
\begin{figure}[h]
	\begin{center}
		\includegraphics[width=0.55\textwidth]{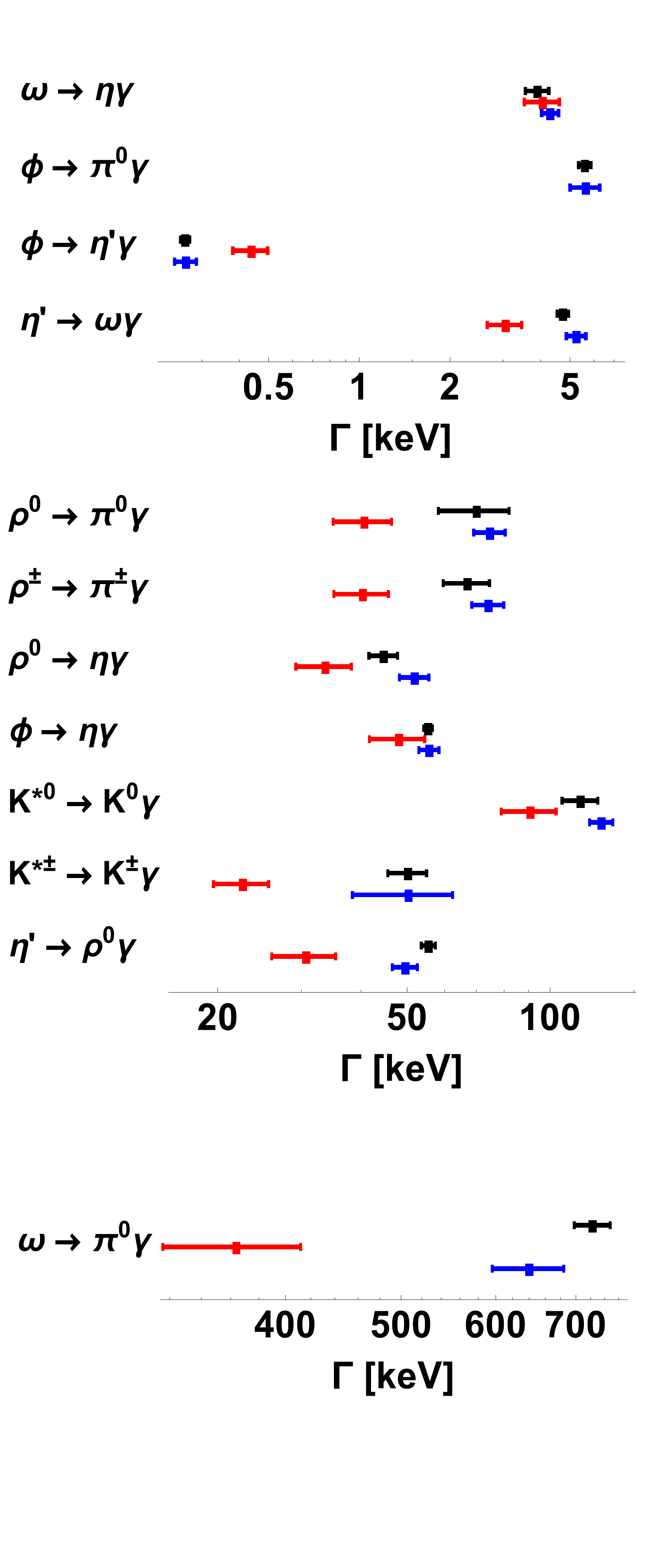}
	 	\caption{Comparison of the decay rates at leading order 
	 	(red, middle entries) and at next-to-leading order in scenario II  (blue, lower entries) with the experimental results (black, upper entries).}
		\label{Abb:VisuLONLO}
	\end{center}
\end{figure}

	\begin{table}[h]
		\renewcommand{\arraystretch}{1.3}
		\begin{center}
			\caption{Decay rates using the two scenarios described in the text. 
			In both cases $\Lambda_1$ is treated as a fit parameter. For the mixing angles we made use of the NLO values $\theta_V=39.8\degree$ and $\theta_P^{[1]}=-12.4\degree$.
			}
			\label{Tab:VariationMatrElem}
			\begin{tabular}{lccc}
				\hline
				\hline
				Decay & 
				$\Gamma _{\mathrm{PDG}}$ (keV) & 
				$\Gamma^\text{I}$ (keV) &
				$\Gamma^\text{II}$ (keV)\\
				\hline
				$\rho ^{0}\rightarrow \pi ^{0}\gamma$ & $70.\pm 12.$ 
				& $70.3 \pm 7.5$ & $74.8 \pm 5.7$ 
				\\ 
				$\rho ^{\pm}\rightarrow \pi ^{\pm}\gamma$ & $67.1 \pm 7.5$ 
				& $69.8 \pm 7.5$ & $74.2 \pm 5.7$ 
				\\ 
				$\rho ^{0}\rightarrow \eta \gamma$ & $44.7\pm 3.1$ 
				& $53.0 \pm 5.5$ & $51.9 \pm 3.7$ 
				\\ 
				$\omega \rightarrow \pi ^{0}\gamma$ & $723.\pm 25.$ 
				& $601. \pm 55.$ & $640. \pm 44.$ 
				\\ 
				$\omega \rightarrow \eta \gamma$ & $3.91\pm 0.35$ 
				& $4.53 \pm 0.46$ & $4.31 \pm 0.28$ 
				\\ 
				$\phi \rightarrow \pi ^{0}\gamma$ & $5.61\pm 0.26$ 
				& $5.61 \pm 0.84$ & $5.65 \pm 0.64$ 
				\\
				$\phi \rightarrow \eta \gamma$ & $55.4\pm 1.1$ 
				& $56.0 \pm 3.6$ & $55.7 \pm 2.7$ 
				\\
				$\phi \rightarrow \eta ' \gamma$ & $0.2643 \pm 0.0090$ 
				& $0.255 \pm 0.029$ & $0.265 \pm 0.022$
				\\ 
				$K^{\ast 0}\rightarrow K^{0}\gamma$ & $116.\pm 10.$ 
				& $133.8 \pm 8.8$ & $128.3 \pm 7.2$ 
				\\ 
				$K^{\ast \pm}\rightarrow K^{\pm}\gamma$ & $50.4\pm 4.6$ 
				& $50. \pm 15.$ & $50. \pm 12.$
				\\ 
				$\eta ' \rightarrow \rho ^{0}\gamma$ & $55.5\pm 1.9$ 
				& $49.4 \pm 4.2$ & $49.6 \pm 3.0$ 
				\\ 
				$\eta ' \rightarrow \omega \gamma$ & $4.74\pm 0.20$ 
				& $5.52 \pm 0.48$ & $5.26 \pm 0.39$
				\\
				\hline
				$\chi ^{2}_{\mathrm{red}}$ & - & 9.2 & 6.1
				\\
				$\Lambda _{1}$ & - & $0.307 \pm 0.074$ & $0.290 \pm 0.050$	
				\\
				\hline
				\hline
			\end{tabular}
		\end{center}
		\renewcommand{\arraystretch}{1}
	\end{table}
	In Figure \ref{Abb:VisuLONLO}, we present a visual comparison of the decay 
rates at leading order (red, middle entries)  and at next-to-leading order in 
scenario II (blue, lower entries) with the experimental results (black, upper entries).
	Here, a clear improvement in the description of the decay rates can be seen 
in the transition from LO to NLO.
	To enable a quantitative comparison, we also show 
the deviations $\delta_1$ and $\delta_2$ of Eqs.~(\ref{defd1d2}) for our 
best-fit results in Table \ref{Tab:ErgebnissePVgammaNLO}.
	Since our calculation is valid up to and including order $1/N_c$ and order 
$\chi$, we expect uncertainties of the order of 
$\sqrt{(1/N_c)^4+(\mathring{M}_K/(4\pi F))^8}$.
	Inserting $N_c=3$ and the values of Table \ref{Tab:WerteGroessenImModell} 
for $\mathring{M}_K$ and $F$, this amounts to relative deviations of the order 
of 12\%.
	After inspecting the column ``deviation $\delta_1$'' of Table 
\ref{Tab:ErgebnissePVgammaNLO}, we find that the relative deviation for almost 
all decays is more or less within this deviation.
	A notable exception is the decay $\rho^0\to\eta\gamma$ with $\delta_1=16$\%.
	The deviations $\delta_2=1.5$, $\delta_2=-1.6$, and $\delta_2=-1.7$ for the
decays $\rho^0\to \eta\gamma$, $\omega\to\pi^0\gamma$, and 
$\eta'\to\rho^0\gamma$, respectively, hint at some tension, 
which will be partially resolved after refining the model.
	The linear combination $c_-=c_5-c_6-c_8$ only enters the charged decay
$K^{\ast\pm}\to K^\pm\gamma$ (see Table \ref{table_Tifull} of Appendix A).
	Therefore, the central values of the experiment and of the fit coincide.
	As a consequence, the remaining linear combination, $c_+=c_5+c_6-c_7$, is 
essentially the only parameter available to describe SU(3)-symmetry-breaking 
effects.

\begin{table}[htb]
	\renewcommand{\arraystretch}{1.3}
	\begin{center}
		\caption{Decay rates using the second scenario described in the text
		together with the deviations $\delta_1$ and $\delta_2$ of 
		Eqs.~(\ref{defd1d2}).}
		\label{Tab:ErgebnissePVgammaNLO}
		\begin{tabular}{lcccc}
			\hline
			\hline
			Decay & $\Gamma_{\mathrm{PDG}}$ (keV) & $\Gamma_{\mathrm{NLO}}$ (keV) & deviation $\delta_{1}$ & deviation $\delta_{2}$
			\\
			\hline
			$\rho^{0}\rightarrow \pi^{0}\gamma$ & $70.\pm 12.$ 
			& $74.8 \pm 5.7$ & $0.066$ & $0.35$
			\\ 
			$\rho^{\pm}\rightarrow \pi^{\pm}\gamma$ & $67.1 \pm 7.5$ 
			& $74.2 \pm 5.7$ & $0.11$ & $0.76$
			\\ 
			$\rho^{0}\rightarrow \eta \gamma$ & $44.7\pm 3.1$ & $51.9 \pm 3.7$ & $0.16$ & $1.5$
			\\ 
			$\omega \rightarrow \pi^{0}\gamma$ & $723.\pm 25.$ & $640. \pm 44.$
			& $-0.11$ & $-1.6$
			\\ 
			$\omega \rightarrow \eta \gamma$ & $3.91\pm 0.35$ & $4.31 \pm 0.28$ 
			& $0.10$ & $0.89$
			\\ 
			$\phi \rightarrow \pi ^{0}\gamma$ & $5.61\pm 0.26$ & $5.65 \pm 0.64$
			& $0.007$ & $0.060$
			\\
			$\phi \rightarrow \eta \gamma$ & $55.4\pm 1.1$ & $55.7 \pm 2.7$ 
			& $0.005$ & $0.093$
			\\
			$\phi \rightarrow \eta ' \gamma$ & $0.2643 \pm 0.0090$ 
			& $0.265 \pm 0.022$ & $-0.002$ & $-0.017$
			\\ 
			$K^{\ast 0}\rightarrow K^{0}\gamma$ & $116.\pm 10.$ 
			& $128.3 \pm 7.2$ & $0.11$ & $1.0$
			\\ 
			$K^{\ast \pm}\rightarrow K^{\pm}\gamma$ & $50.4\pm 4.6$ 
			& $50. \pm 12.$ & - & -
			\\ 
			$\eta ' \rightarrow \rho ^{0}\gamma$ & $55.5\pm 1.9$ 
			& $49.6 \pm 3.0$ & $-0.11$ & $-1.7$
			\\ 
			$\eta ' \rightarrow \omega \gamma$ & $4.74\pm 0.20$ 
			& $5.26 \pm 0.39$ & $0.11$ & $1.2$
			\\
			\hline
			\hline
		\end{tabular}
	\end{center}
	\renewcommand{\arraystretch}{1}
\end{table}

   In Table \ref{table_correlation_coefficients}, we present the correlation 
coefficients $\langle\delta c_i\delta c_j\rangle/(\delta c_i\delta c_j)$ for
our best fit of Table \ref{Tab:ErgebnissePVgammaNLO}.
	As one might expect, the strongest correlation exists between parameters 
$c_1$ and $c_+$, because the linear combination $\tilde{c}_1=c_1+2M_\pi^2c_+$ 
contributes to all decays.
	There is also an equally strong correlation between $c_1$ and $c_2$.
	The parameter $c_2$ only contributes to the transitions between the 
vector-meson singlet and the pseudoscalar-meson octet. 
	There is a slightly smaller correlation between $c_2$ and $c_+$.
	Finally, the last notable correlations exist between the parameter 
$\Lambda_1$, which is of order $1/N_c$, and the parameters $c_1$ and $c_3$.
	The remaining correlations are negligibly small.

\renewcommand{\arraystretch}{1.3}
\begin{table}[ht]
	\caption{Off-diagonal array containing the correlation coefficients 
	$\langle\delta c_i\delta c_j\rangle/(\delta c_i\delta c_j)$
	of the parameters $c_i$ for the fit of Table \ref{Tab:ErgebnissePVgammaNLO}.}
	\label{table_correlation_coefficients}
	\begin{tabular}{l|ccccc}
		\hline
		\hline
		$c_2$&$\quad-0.73$&&&&\\
		$c_3$&$\quad0.033$&\quad$0.14$&&&\\
		$c_+$&$\quad-0.78$&\quad$0.46$&\quad$0.16$&&\\
		$c_-$&$\quad 0.21$&\quad$-0.16$&\quad$0.020$&\quad$-0.13$&\\
		$\Lambda_{1}$&$\quad 0.50$ &\quad$-0.19$ &\quad $0.50$ &\quad $-0.29$ 
		&\quad $0.11$\\
		\hline
		&\quad$c_1$&\quad$c_2$&\quad$c_3$&\quad$c_+$ &\quad$c_{-}$\\
		\hline
		\hline
	\end{tabular}
\end{table}
\renewcommand{\arraystretch}{1}

\subsection{Expansion of the quark-charge matrix in $1/N_c$}
\begin{figure}
	\begin{center}
		\includegraphics[width=0.55\textwidth]{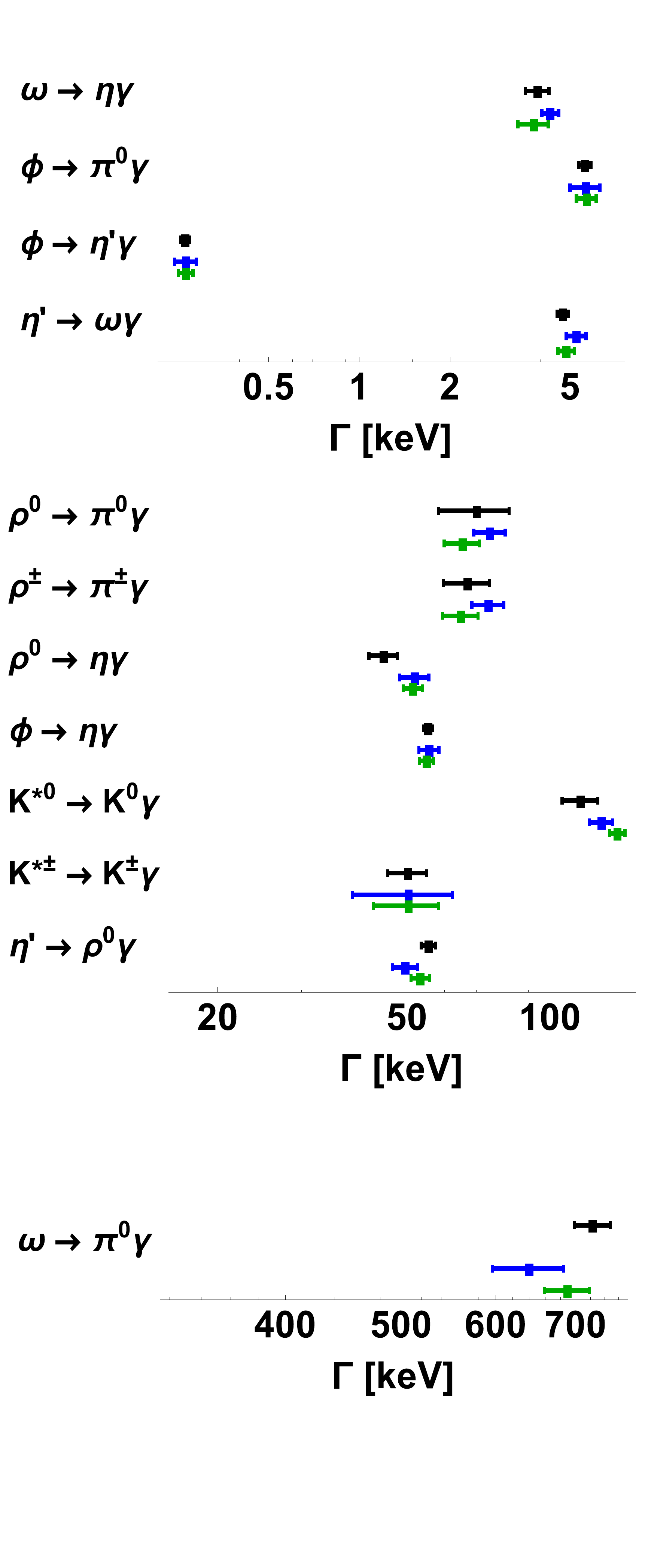}
		\caption{Comparison of the decay rates at next-to-leading order in 
		scenario II (blue, middle entries),
		next-to-leading order including a $1/N_c$ expansion of the quark-charge matrix (green, lower entries) with the experimental results (black, upper entries).}
		\label{Abb:VisuNLOQphysQNc}
	\end{center}
\end{figure}

   As our final example, we also include the expansion of the quark-charge 
matrix in 1/$N_c$ [see Eq.~(\ref{Q})].
   As a consequence of this expansion, also the $c_4$ interaction Lagrangian of Eq.~(\ref{LagrangianNLOPVgammaLNc}) contributes
to the invariant amplitudes (see Table \ref{table_TiLNc} of Appendix 
\ref{appendix_Lagrangians}).
   Using the expressions of Table \ref{table_TiLNc} of Appendix
\ref{appendix_Lagrangians} and applying scenario II
of Sec.~\ref{subsection_LNc_quark_mass}, we obtain the results shown in Table \ref{table_fit_Qexp}.
   In fact, this scenario provides us with one additional parameter and it is 
therefore not surprising that $\chi^2_\text{red}=2.8$ (5 degrees of freedom) is 
smaller than the corresponding value $\chi^2_\text{red}=6.6$ of Table \ref{Tab:VariationMatrElem}.
	The parameters of the fit are given by
\begin{equation}
	\begin{split}
	c_1&=0.0536 \pm 0.0013,\quad 
	c_2=-0.000613 \pm 0.000078,\quad
	c_3=0.00109 \pm 0.00027,\\
	c_4&=0.00142 \pm 0.00055,\quad
	c_+=\left(1.05 \pm 0.45 \right)\cdot 10^{-9}\,\mathrm{MeV}^{-2},\quad
	c_-=\left(5.1 \pm 1.6 \right)\cdot 10^{-9}\,\mathrm{MeV}^{-2},\\
	\Lambda_1&= 0.247 \pm 0.032.
	\end{split}
\end{equation}
	In Figure \ref{Abb:VisuNLOQphysQNc}, we present a visual comparison of the 
decay rates at NLO in scenario II (blue, middle entries) 
and at NLO including a quark-charge expansion (green, lower entries) with the 
experimental results (black, upper entries).
	Except for the decays $\rho^0\to\eta\gamma$ and $K^{\ast0}\to K^0\gamma$,
we obtain an excellent agreement between experiment and theory. 
	
\renewcommand{\arraystretch}{1.3}
\begin{table}[htb]
	\begin{center}
		\caption{Decay rates using the second scenario described in the text, 
			including an expansion of the quark-charge matrix in $1/N_c$.
			For the mixing angles we made use of the NLO values
			$\theta_V=39.8\degree$ and $\theta_P^{[1]}=-12.4\degree$. $\chi ^{2}_{\mathrm{red}}=2.8$ (5 degrees of freedom).}
		\label{table_fit_Qexp}
		\begin{tabular}{lcccc}
			\hline
			\hline
			Decay & $\Gamma_{\mathrm{PDG}}$ (keV) & $\Gamma_{\mathrm{mod}}$ in keV & $\delta_{1}$ & $\delta_{2}$
			\\
			\hline
			$\rho ^{0}\rightarrow \pi ^{0}\gamma$ & $70.\pm 12.$ 
			& $65.5 \pm 5.6$ & $-0.066$ & $-0.35$
			\\ 
			$\rho ^{\pm}\rightarrow \pi ^{\pm}\gamma$ & $67.1 \pm 7.5$ 
			& $65.0 \pm 5.6$ & $-0.031$ & $-0.23$
			\\ 
			$\rho ^{0}\rightarrow \eta \gamma$ & $44.7\pm 3.1$ 
			& $51.5 \pm 2.4$ & $0.15$ & $1.7$
			\\ 
			$\omega \rightarrow \pi ^{0}\gamma$ & $723.\pm 25.$ 
			& $689. \pm 30.$ & $-0.047$ & $-0.87$
			\\ 
			$\omega \rightarrow \eta \gamma$ & $3.91\pm 0.35$ 
			& $3.80 \pm 0.44$ & $-0.028$ & $-0.19$
			\\ 
			$\phi \rightarrow \pi ^{0}\gamma$ & $5.61\pm 0.26$ 
			& $5.69 \pm 0.43$ & $0.014$ & $0.16$
			\\
			$\phi \rightarrow \eta \gamma$ & $55.4\pm 1.1$ & $55.0 \pm 1.8$ 
			& $-0.0070$ & $-0.18$
			\\
			$\phi \rightarrow \eta ' \gamma$ & $0.2643 \pm 0.0090$ 
			& $0.266 \pm 0.015$ & $0.0058$ & $0.090$
			\\ 
			$K^{\ast 0}\rightarrow K^{0}\gamma$ & $116.\pm 10.$ 
			& $138.6 \pm 5.2$ & $0.19$ & $2.0$
			\\ 
			$K^{\ast \pm}\rightarrow K^{\pm}\gamma$ & $50.4\pm 4.6$ 
			& $50.4 \pm 7.9$ & - & -
			\\ 
			$\eta ' \rightarrow \rho ^{0}\gamma$ & $55.5\pm 1.9$ 
			& $53.4 \pm 2.4$ & $-0.038$ & $-0.69$
			\\ 
			$\eta ' \rightarrow \omega \gamma$ & $4.74\pm 0.20$ 
			& $4.87 \pm 0.31$ & $0.027$ & $0.34$
			\\
			\hline
			\hline
		\end{tabular}
	\end{center}
\end{table}
\renewcommand{\arraystretch}{1}

\subsection{Coupling constants and convergence}
	We have organized the Lagrangians in terms of $1/N_c$ and the quark masses 
$m$ (contained in the quantities $\chi_\pm$).
	For the number of colors we insert $N_c=3$ and, with respect to the 
quark-mass expansion, we consider $M_K^2/(4\pi F)^2\approx1/4$ as a typical 
small dimensionless expansion parameter, where $\Lambda_\chi=4\pi F$ denotes the chiral-symmetry-breaking scale \cite{Manohar:1983md}.
	In Table \ref{table_coupling_constants}, we collect the coupling constants 
as obtained from fitting the data using different levels of approximation.
    The second column (LO) refers to the leading-order Lagrangian with 
$\theta_P=\theta_P^{[0]}=-19.7\degree$ and ideal
$\phi$-$\omega$ mixing, the third column (LO+$1/N_c$) to the leading-order 
Lagrangian plus $1/N_c$ corrections with $\theta_P=\theta_P^{[0]}=-19.7\degree$ and ideal $\phi$-$\omega$ mixing, the fourth column to the complete next-to-leading-order Lagrangian without expanding the quark-charge matrix $Q$ (NLO), and the fifth column to the complete next-to-leading-order Lagrangian including an expansion of $Q$ (NLO, $Q$ expanded).
	The last two scenarios made use of
$\theta_P=\theta_P^{[1]}=-12.4\degree$, $\theta_V=39.8\degree$, and physical 
values for $F_\pi$ and $F_K$.
   As can be seen by comparing Tables \ref{table_Tifull} and \ref{table_TiLNc} 
of Appendix \ref{appendix_Lagrangians}, the contributions of the coefficients 
$c_i$ to the decay matrix elements are redistributed in the version including the expansion of the quark-charge matrix.
   This is then the reason why, except for $c_1$, the coefficients differ 
notably for the last two cases.

\renewcommand{\arraystretch}{1.3}
\begin{table}[ht]
	\caption{Coupling constants determined at leading order (LO), leading order
	plus $1/N_c$ corrections, next-to-leading order (NLO), and next-to-leading order with expanded quark-charge matrix
	(NLO, $Q$ expanded). See text for details.}
	\label{table_coupling_constants}
	\begin{tabular}{lcccc}
		\hline
		\hline
		Coupling constant \quad&\quad LO\quad &\quad LO+$1/N_c$ \quad 
		&\quad NLO \quad&\quad NLO, $Q$ expanded
		\\
		\hline
		$c_{1}$ $\left[ 10^{-2} \right]$ \quad&\quad $3.82 \pm 0.25$ \quad
		&\quad $3.36 \pm 0.20$ \quad&\quad $5.22 \pm 0.20$ \quad&\quad $5.36 \pm 0.13$
		\\
		$c_{2}$ $\left[ 10^{-2} \right]$ \quad&\quad - \quad
		&\quad $0.67 \pm 0.10$ \quad&\quad $-0.100 \pm 0.021$ \quad&\quad $-0.0613 \pm 0.0078$
		\\
		$c_{3}$ $\left[ 10^{-2} \right]$ \quad&\quad - \quad
		&\quad $-0.39 \pm 0.25$ \quad&\quad $0.272 \pm 0.072$ \quad&\quad $0.109 \pm   0.027$
		\\
		$c_{4}$ $\left[ 10^{-2} \right]$ \quad&\quad - \quad&\quad - \quad
		&\quad - \quad&\quad $0.142 \pm 0.055$
		\\
		$c_{+}$ $\left[ 10^{-2}\,\mathrm{GeV}^{-2}\right]$ \quad&\quad - \quad
		&\quad - \quad&\quad $0.280 \pm  0.078$ \quad&\quad $0.105 \pm 0.045$
		\\
		$c_{-}$ $\left[ 10^{-2}\,\mathrm{GeV}^{-2} \right]$ \quad&\quad - \quad
		&\quad - \quad&\quad $-0.61 \pm 0.39$ \quad&\quad $0.51 \pm 0.16$
		\\
		\hline
		\hline
	\end{tabular}
\end{table}
\renewcommand{\arraystretch}{1}

   Finally, we would like to comment on the order of magnitude of the 
corrections in comparison with the leading-order term.
	We multiply the constants $c_2$, $c_3$, and $c_4$ by a factor of 3 to 
obtain the coefficients belonging to the $1/N_c$ expansion.
	Similarly, we multiply $c_+$ and $c_-$ by $(4\pi F)^2$ to obtain the 
coefficients for the dimensionless quark-mass expansion.
	The results corresponding to the last two columns of Table 
\ref{table_coupling_constants} are shown in Table
\ref{table_convergence_coupling_constants}.
\renewcommand{\arraystretch}{1.3}
\begin{table}[ht]
	\caption{Expansion coefficients at next-to-leading order (NLO) 
	corresponding to the last two columns of Table \ref{table_coupling_constants}. For simplicity we suppress uncertainties.}
	\label{table_convergence_coupling_constants}
	\begin{tabular}{lcc}
		\hline
		\hline
		Coefficients \quad&\quad physical $Q$ \quad&\quad $Q$ expanded
		\\
		\hline
		$\tilde{c}_{1}$ [$10^{-2}$]\quad&\quad $5.23$ \quad&\quad $5.36$
		\\
		$c_{2} \cdot N_{c}$ [$10^{-2}$]\quad&\quad $-0.30$ \quad&\quad $-0.18$
		\\
		$c_{3} \cdot N_{c}$ [$10^{-2}$]\quad&\quad $0.82$ \quad&\quad $0.33$
		\\
		$c_{4} \cdot N_{c}$ [$10^{-2}$]\quad&\quad - \quad&\quad $0.43$
		\\
		$c_{+} \cdot \left( 4\pi F \right)^{2}$ [$10^{-2}$]\quad&\quad $0.36$ 
		\quad&\quad $0.14$
		\\
		$c_{-} \cdot \left( 4\pi F \right)^{2}$ [$10^{-2}$]\quad&
		\quad $-0.79$ \quad&\quad $0.67$
		\\
		\hline
		\hline
	\end{tabular}
\end{table}
\renewcommand{\arraystretch}{1}
	Let us have a closer look at the implications of the second column of Table 
\ref{table_convergence_coupling_constants}	(NLO with physical quark-charge 
matrix $Q$).
	We notice that all of the amplitudes ${\cal A}_i$ of 
Eq.~(\ref{leffVPgammaappendix}) except for ${\cal A}_5$ start with $c_1$.
	We multiply each ${\cal A}_i$ ($i\neq 5$) with a suitable factor such that 
the leading-order term is simply given by $c_1$.
	We can then easily identify the amount of the largest relative correction.
	Regarding the $1/N_c$ terms, this is $3c_2/(2c_1)$ for the amplitudes 
${\cal A}_9$ and ${\cal A}_{11}$ and $3 c_3/(2c_1)$ for the amplitudes 
${\cal A}_7$ and ${\cal A}_{10}$, respectively.
	Using the values of the second column of Table 
\ref{table_convergence_coupling_constants}, we obtain $-0.086$
and $0.23$, respectively, where we have neglected the uncertainties.
	Keeping in mind that these numbers still have to be multiplied by 1/3, the 
$1/N_c$ corrections turn out to	be relatively small, namely $-2.9$\% and
7.8\%, respectively.
	For the quark-mass corrections, the largest correction originating from 
$c_+$ is found in the ${\cal A}_4$ amplitude, namely, the ratio 
$16|c_+|(4\pi F)^2/(3c_1)=0.37$ which gets multiplied by 
$(M_K^2-M_\pi^2)/(4\pi F)^2=0.17$.
	The relative quark-mass correction of 6.3\% is of a similar magnitude as 
the $1/N_c$ correction.
	More pronounced is the case of the $c_-$ coupling, resulting in the ratio
$6|c_-|(4\pi F)^2/c_1=0.90$ which, together with the factor 
$(M_K^2-M_\pi^2)/(4\pi F)^2=0.17$, gives rise to a relative correction of 15\%.
	Recall that this parameter is entirely determined by the decay 
$K^{\ast\pm}\to K^\pm\gamma$.
	For the third column of Table \ref{table_convergence_coupling_constants}
(NLO with expanded quark-charge matrix $Q$), we obtain similar results.

\subsection{Comparison with other calculations in chiral effective Lagrangian 
approaches}

	Reference \cite{Klingl:1996by} contains the leading-order Lagrangian of the 
vector formulation for the $VP\gamma$ decay of neutral vector mesons into 
neutral pions.
	When comparing this with Eq.~(3.19) of Ref.~\cite{Klingl:1996by}, we agree 
after identifying our $2e c_1/F$ with $d/f_\pi$ of Ref.~\cite{Klingl:1996by}.\footnote{Note that the vector-meson matrix of Ref.~\cite{Klingl:1996by} is two times our vector-meson matrix.}
	However, their Eq.~(4.7) for the decay rates seems to contain an error, 
namely, the second line needs to be multiplied by a factor of 1/9, originating 
from the elements of the quark-charge matrix in the form $(2/3-1/3)^2$.
	Accordingly, the coupling $d\simeq 0.01$ of Eq.~(4.10) needs to be
multiplied by a factor of 3.
	Similarly, our results from the leading-order Lagrangian agree with the 
coefficients reported in Fig.~3 of Ref.~\cite{Danilkin:2017lyn}, which, beyond 
SU(3) symmetry, implicitly made use of nonet symmetry in combination with ideal $\phi$-$\omega$ mixing and neglected $\eta$-$\eta'$ mixing.

   	The antisymmetric tensor-field representation 
\cite{Kyriakopoulos:1969zm,Kyriakopoulos:1973pt} was used in, e.g.,
Refs.~\cite{RuizFemenia:2003hm,Lutz:2008km,Terschlusen:2012xw,Chen:2013nna} for 
the calculation of the $VP\gamma$ interaction.
   	In Ref.~\cite{RuizFemenia:2003hm}, the relevant interaction Lagrangian for 
the interaction of two vector fields with one pseudoscalar field ($VVP$) and for one vector resonance with an external vector field and a pseudoscalar field ($VJP$) was 
constructed, involving $7+4$ coupling constants, respectively.
	In terms of the QCD short-distance behavior of the $VVP$ Green function, 
constraints among the coupling constants were derived.
   	Using these constraints, Eq.~(4.2) of Ref.~\cite{RuizFemenia:2003hm} 
provides a parameter-free prediction for the $\omega\to\pi^0\gamma$-transition 
matrix element, translating into a prediction for our $c_1$,
\begin{displaymath}
|c_1|=\frac{1}{4\sqrt{2}}\left(\frac{3}{8\pi^2}\frac{m_\omega}{F}
-\frac{F}{2}\frac{m_\omega}{m_V^2}\right).
\end{displaymath}
   	Using $F=90.9$~MeV and $m_\omega=m_V=782.7$~MeV, one obtains 
$|c_1|=4.76\times 10^{-2}$ which has to be compared with our LO prediction 
$|c_1|=(3.82\pm 0.25)\times 10^{-2}$ and the NLO prediction
$|c_1|=(5.22\pm 0.20)\times 10^{-2}$ of Table \ref{table_coupling_constants}.
   	In Ref.~\cite{Lutz:2008km}, antisymmetric tensor fields were used for 
describing the radiative decays of the vector-meson nonet into the pseudoscalar 
octet.
   	The $\eta'$ was not considered, the physical $\eta$ was taken as part of 
the pseudoscalar octet, and for the $\phi$-$\omega$ system an ideal mixing was 
assumed.
   	The decay proceeds either via a $VVP$ vertex such that the propagating 
neutral vector meson subsequently couples to a real photon or via a direct 
$VP\gamma$ interaction (which is considered to be of higher order in
their chiral counting).
   	The decay rates then contain three (combinations of) coupling constants, 
namely $e_A$ (direct decay), $h_A e_V$ and $b_A e_V$ (indirect decay) 
[see Eqs.~(38)-(42) of Ref.~\cite{Lutz:2008km}].
   	In the limit of SU(3) symmetry, our results for the invariant amplitudes 
fully agree with those of Ref.~\cite{Lutz:2008km}.
   	To see this, one needs to set all vector-meson masses equal to $m_V$, all 
pseudoscalar meson masses equal to $\overline{M}$, $F=f$, and, finally,
$e|\tilde{c}_1|=|\tilde{e}_A|/8$, where 
$\tilde{e}_A=e_A+\frac{1}{4}h_Ae_V-2b_Ae_V \overline{M}^2/m_V^2$.
   	In Ref.~\cite{Terschlusen:2012xw}, the analysis was extended to also 
include the $\eta'$ meson.
   	With two additional parameters, namely, the $\eta$-$\eta'$ mixing angle 
$\theta_P$ and one parameter $b_H$ for the interaction of the singlet eta with 
two vector mesons, in total five parameters were adjusted to five decays.
   	In particular, an unconventionally small mixing angle $\theta_P\simeq\pm 
2\degree$ was found.
   	When taking SU(3)-symmetry breaking effects into account, in our framework 
two additional parameters are available, namely, $c_+$ and $c_-$, whereas in 
the framework of Ref.~\cite{Lutz:2008km} only the combination $b_A e_V$ will give rise to SU(3)-symmetry breaking effects.
   	This term corresponds to our $c_+$ structure.
   	In particular, the SU(3) relation $|{\cal A}_{K^{\ast 0}\to 
K^0\gamma}|=2|{\cal A}_{K^{\ast\pm}\to K^\pm\gamma}|$ will not be broken in 
the framework of Ref.~\cite{Lutz:2008km}, unless higher-order terms are taken into account.
   	In our calculation, the parameter $c_-$ decouples ${\cal A}_{K^{\ast\pm}\to 
K^\pm\gamma}$ and ${\cal A}_{K^{\ast 0}\to K^0\gamma}$.
   	On the other hand, in Ref.~\cite{Chen:2013nna}, the importance of such a 
term in the context of SU(3) symmetry breaking was already worked out for the 
radiative $K^\ast\to K\gamma$ decays.
   	Reference~\cite{Chen:2013nna} extended the results of 
\cite{RuizFemenia:2003hm} by also including excited vector-meson
resonances.

	Recently, Kimura, Morozumi, and Umeeda \cite{Kimura:2016xnx} investigated 
decays of light hadrons within a chiral Lagrangian model that includes both 
the lightest pseudoscalar and vector mesons.
	As an extension of chiral perturbation theory, they included one-loop 
corrections due to the Goldstone bosons and the corresponding counter terms.
	In addtion to other processes, they also particularly looked at the 
$VP\gamma$ reactions.
	For the decays involving the light vector mesons, the model provides three 
	parameters.
	
	In Table \ref{Tab:VergleichModErgebnisse}, we provide a comparison of our 
results with those of Ref.~\cite{Kimura:2016xnx}.
	First of all, we note that, with the exception of the decays 
$\omega\to\pi^0\gamma$ and $K^{\ast0}\to K^0\gamma$, our central values for 
the decay rates agree better with the experimental values than those from 
Ref.~\cite{Kimura:2016xnx}.
	In addition, the uncertainties from Ref.~\cite{Kimura:2016xnx} are, on 
average, much larger than ours.
	We conclude that our approach provides an improved description of the 
$VP\gamma$ decays. 

\renewcommand{\arraystretch}{1.3}
\begin{table}[htb]
	\begin{center}
		\caption{Comparison of our results at NLO with the predictions of 
		Ref.~\cite{Kimura:2016xnx} (KMU18).}
		\label{Tab:VergleichModErgebnisse}
		\begin{tabular}{lccc}
			\hline
			\hline
			Decay & $\Gamma_{\mathrm{exp}}$ [keV] & $\Gamma _{\mathrm{NLO}}$ [keV] & $\Gamma_{\mathrm{KMU18}}$ [keV]
			\\
			\hline
			$\rho ^{0}\rightarrow \pi ^{0}\gamma$ & $70.\pm 12.$ 
			& $65.5 \pm 5.6$ & $46. \pm 5.$
			\\ 
			$\rho ^{\pm}\rightarrow \pi ^{\pm}\gamma$ & $67.1 \pm 7.5$ 
			& $65.0 \pm 5.6$ & $73. \pm 7.$
			\\ 
			$\rho ^{0}\rightarrow \eta \gamma$ & $44.7\pm 3.1$ 
			& $51.5 \pm 2.4$ & $33.^{+8.}_{-9.}$
			\\ 
			$\omega \rightarrow \pi ^{0}\gamma$ & $723.\pm 25.$ 
			& $689. \pm 30.$ & $710. \pm 90.$
			\\ 
			$\omega \rightarrow \eta \gamma$ & $3.91\pm 0.35$ 
			& $3.80 \pm 0.44$ & $5.5^{+1.6}_{-1.3}$
			\\ 
			$\phi \rightarrow \pi ^{0}\gamma$ & $5.61\pm 0.26$ 
			& $5.69 \pm 0.43$ & $17.^{+12.}_{-9.}$
			\\
			$\phi \rightarrow \eta \gamma$ & $55.4\pm 1.1$ & $55.0 \pm 1.8$ 
			& $22.^{+9.}_{-12.}$
			\\
			$\phi \rightarrow \eta ' \gamma$ & $0.2643 \pm 0.0090$ 
			& $0.266 \pm 0.015$ & $0.39^{+0.12}_{-0.09}$
			\\ 
			$K^{\ast 0}\rightarrow K^{0}\gamma$ & $116.\pm 10.$ 
			& $138.6 \pm 5.2$ & $110. \pm 10.$
			\\ 
			$K^{\ast \pm}\rightarrow K^{\pm}\gamma$ & $50.4\pm 4.6$ 
			& $50.4 \pm 7.9$ & $28. \pm 3.$
			\\ 
			$\eta ' \rightarrow \rho ^{0}\gamma$ & $55.5\pm 1.9$ 
			& $53.4 \pm 2.4$ & -
			\\ 
			$\eta ' \rightarrow \omega \gamma$ & $4.74\pm 0.20$ 
			& $4.87 \pm 0.31$ & $4.6^{+3.3}_{-2.0}$
			\\
			\hline
			\hline
		\end{tabular}
	\end{center}
\end{table}
\renewcommand{\arraystretch}{1}

\section{summary}
	In this work we analyzed the radiative transitions between the vector-meson 
nonet and the pseudoscalar-meson nonet within a chiral effective Lagrangian 
approach.
	For that purpose we have determined the Lagrangian up to and including the 
next-to-leading order in an expansion in $1/N_c$ and the quark masses. 
	For the transformation behavior of the vector mesons under the Lorentz
group we made use of the vector representation.
	At leading order, the Lagrangian contains one free parameter which was 
determined from a simultaneous fit to 11 experimental decay rates. (The decay
$\phi\to\pi^0\gamma$ was excluded because of the OZI rule.)
	Both $\eta$-$\eta'$ and $\phi$-$\omega$ mixing were taken into account at 
leading order.  
	The results of this scenario are given in Table \ref{table_fit_1} and 
clearly show that a good description of all experimental data at leading order 
is not possible.
	We then gradually improved the model by first taking into account $1/N_c$ 
corrections (see Table \ref{table_fit_SU(3)}) and then also quark mass 
corrections (see Table \ref{Tab:VariationConstantLambda1}), which introduced 2 plus 2 additional parameters, respectively.
	In this context, we also had to consider the corrections due to the 
wavefunction renormalization and the mixing at NLO.
	As a result, another parameter of the $\eta$-$\eta'$ system was included in 
the calculation, which served as a further fit parameter.
	Then fits were carried out in which either the invariant amplitude or the 
decay were expanded up to and including next-to-leading order 
(scenario I and II, respectively).
	From Table \ref{Tab:VariationMatrElem} we concluded that the second 
scenario yields a better description of the experimental data.
	In Figure \ref{Abb:VisuNLOQphysQNc} we provided a visual presentation of 
the improvement from LO to NLO.
	However, one has to keep in mind in this context that at leading order only 
one free parameter is available to describe 11 decays, while at NLO 6 
parameters (including $\Lambda_1$) have been fitted to 12 decays.
	In our final fit we made use of a $1/N_c$ expansion of the quark-charge 
matrix, which gives rise to one additional free parameter and results in the
best description of the data (see Table \ref{table_fit_Qexp} and 
Fig.~\ref{Abb:VisuNLOQphysQNc}).
	We also found that the contributions to the amplitudes
${\cal A}_i$ of Eq.~(\ref{leffVPgammaappendix}), which are generated by the 
$1/N_c$ and the quark mass corrections, are smaller in magnitude than 15\% of the leading-order term.
	In other words, they can really be regarded as corrections.
	Clearly, our approach is limited in the sense that extending it to include 
even NNLO corrections would introduce a number of additional parameters such 
that there would be no more predictive power.
	However, it would be interesting to see how pseudoscalar-meson loop 
corrections affect our findings \cite{Krasniqi}.
	In summary, we can say that the present calculation is the most
comprehensive investigation of the $VP\gamma$ decays to date and provides a 
satisfactory description of the experimental decay rates.

\begin{acknowledgements}
   Supported by the Deutsche Forschungsgemeinschaft DFG through the 
Collaborative Research Center
``The Low-Energy Frontier of the Standard Model" (SFB 1044).
   The authors would like to thank Wolfgang Gradl for discussions of the 
experimental results.
\end{acknowledgements}

\begin{appendix}

\section{Lagrangians}
\label{appendix_Lagrangians}

   Let us introduce the following structures involving the singlet and octet 
fields:
\begin{align}
\label{structures_Ti}
T_{1\rho\sigma}&=\rho^+_\rho\partial_\sigma\pi^-+\rho^-_\rho\partial_\sigma\pi^+
+\rho^0_\rho\partial_\sigma\pi^0,\nonumber\\
T_{2\rho\sigma}&=K^{\ast+}_\rho\partial_\sigma K^-+K^{\ast-}_\rho\partial_\sigma K^+,\nonumber\\
T_{3\rho\sigma}&=\overline{K}^{\ast 0}_\rho\partial_\sigma K^0+K^{\ast0}_\rho\partial_\sigma\overline{K}^0,\nonumber\\
T_{4\rho\sigma}&=\omega_{8\rho}\partial_\sigma\eta_8,\nonumber\\
T_{5\rho\sigma}&=\omega_{1\rho}\partial_\sigma\eta_1,\nonumber\\
T_{6\rho\sigma}&=\rho^0_\rho\partial_\sigma\eta_8,\nonumber\\
T_{7\rho\sigma}&=\rho^0_\rho\partial_\sigma\eta_1,\nonumber\\
T_{8\rho\sigma}&=\omega_{8\rho}\partial_\sigma\pi^0,\nonumber\\
T_{9\rho\sigma}&=\omega_{1\rho}\partial_\sigma\pi^0,\nonumber\\
T_{10\rho\sigma}&=\omega_{8\rho}\partial_\sigma\eta_1,\nonumber\\
T_{11\rho\sigma}&=\omega_{1\rho}\partial_\sigma\eta_8.
\end{align}
   The effective Lagrangian of the $VP\gamma$ interaction may then be written as
\begin{equation}
\label{leffVPgammaappendix}
{\cal L}_{\rm eff}^{VP\gamma}=e\epsilon^{\mu\nu\rho\sigma}F_{\mu\nu}
\sum_{i=1}^{11}{\cal A}_i T_{i\rho\sigma}.
\end{equation}
   Defining
\begin{align*}
E^{\rho\sigma}&\equiv 2\frac{e}{F}\epsilon^{\rho\sigma\mu\nu} F_{\mu\nu},\\
H_{\rho\sigma}&\equiv\Big[(Q_u+Q_d)T_{1\rho\sigma}+(Q_u+Q_s)T_{2\rho\sigma}
+(Q_d+Q_s)T_{3\rho\sigma}\\
&\quad+\frac{1}{3}(Q_u+Q_d+4Q_s)T_{4\rho\sigma}
+\frac{2}{3}(Q_u+Q_d+Q_s)T_{5\rho\sigma}\\
&\quad+(Q_u-Q_d)\left(\frac{1}{\sqrt{3}}T_{6\rho\sigma}
+\sqrt{\frac{2}{3}}T_{7\rho\sigma}
+\frac{1}{\sqrt{3}}T_{8\rho\sigma}
+\sqrt{\frac{2}{3}}T_{9\rho\sigma}\right)\\
&\quad+\frac{\sqrt{2}}{3}(Q_u+Q_d-2Q_s)(T_{10\rho\sigma}+T_{11\rho\sigma})
\Big],
\end{align*}
where $Q_u$, $Q_d$, and $Q_s$ denote the quark charges,
the Lagrangian of Eq.~(\ref{LagrangianLOPVgamma}) is given by
\begin{equation}
\label{LagLOT}
{\cal L}^{VP\gamma}_{\text{LO}}=c_1 E^{\rho\sigma}H_{\rho\sigma}.
\end{equation}
   The charge factors are given in Table \ref{table_charge_factors} both for
the physical charges and their values as $N_c\to\infty$.
   Note that the $T_{5\rho\sigma}$ term does not contribute for physical values
of the quark charges.
\renewcommand{\arraystretch}{1.3}
\begin{table}[h]
\caption{Charge factors for physical values and the limit $N_c\to\infty$.}
\label{table_charge_factors}
\begin{center}
\begin{tabular}{l r r}
\hline
\hline
{}\quad&\quad Physical value \quad&\quad value as $N_c\to\infty$\\
\hline
$Q_u$\quad&\quad$\frac{2}{3}$\quad&\quad$\frac{1}{2}$\\
$Q_d$\quad&\quad$-\frac{1}{3}$\quad&\quad$-\frac{1}{2}$\\
$Q_s$\quad&\quad$-\frac{1}{3}$\quad&\quad$-\frac{1}{2}$\\
$Q_u+Q_d$\quad&\quad$\frac{1}{3}$\quad&\quad$0$\\
$Q_u+Q_s$\quad&\quad$\frac{1}{3}$\quad&\quad$0$\\
$Q_d+Q_s$\quad&\quad$-\frac{2}{3}$\quad&\quad$-1$\\
$Q_u-Q_d$\quad&\quad$1$\quad&\quad$1$\\
$Q_u+Q_d+4Q_s$\quad&\quad$-1$\quad&\quad$-2$\\
$Q_u+Q_d-2Q_s$\quad&\quad$1$\quad&\quad$1$\\
$Q_u+Q_d+Q_s$\quad&\quad$0$\quad&\quad$-\frac{1}{2}$\\
\hline
\hline
\end{tabular}
\end{center}
\end{table}
\renewcommand{\arraystretch}{1}

   The $1/N_c$ corrections are given by
\begin{align}
\label{LagrangianNLOPVgammaexpapp}
{\cal L}^{VP\gamma}_{\text{NLO,$1/N_c$}}&=E^{\rho\sigma}
\left\{c_2\left[(Q_u+Q_d+Q_s)T_{5\rho\sigma}
+\sqrt{\frac{3}{2}}(Q_u-Q_d)T_{9\rho\sigma}
+\frac{1}{\sqrt{2}}(Q_u+Q_d-2Q_s)T_{11\rho\sigma}\right]\right.\nonumber\\
&\quad+c_3\left[
(Q_u+Q_d+Q_s)T_{5\rho\sigma}
+\sqrt{\frac{3}{2}}(Q_u-Q_d)T_{7\rho\sigma}
+\frac{1}{\sqrt{2}}(Q_u+Q_d-2Q_s)T_{10\rho\sigma}\right]\nonumber\\
&\left.\vphantom{\sqrt{\frac{3}{2}}}\quad
+c_4(Q_u+Q_d+Q_s)(T_{1\rho\sigma}+T_{2\rho\sigma}+T_{3\rho\sigma}
+T_{4\rho\sigma}+T_{5\rho\sigma})
\right\}.
\end{align}
   Since $\langle Q\rangle=Q_u+Q_d+Q_s=0$ for the physical quark charges,
the last term does not contribute in this case.
   Furthermore, for physical quark charges, there is no singlet-to-singlet 
transition ($T_{5\rho\sigma}$).
   In order to express the quark mass corrections we make use of the 
leading-order kaon and pion masses squared \cite{Gasser:1984gg},
$\mathring{M}^2_K= B_0\left(\hat m+m_s\right)$ and 
$\mathring{M}^2_\pi=2 B_0\hat m$.
   To the order we are considering, we replace the leading-order expressions by 
the physical values, i.e., $\mathring{M}^2_K\to M_K^2$ and 
$\mathring{M}^2_{\pi}\to M_\pi^2$.
   The quark-mass corrections are then given by
\begin{align*}
{\cal L}^{VP\gamma}_{\text{NLO,$\chi$}}&=(2c_5-c_7-c_8) E^{\rho\sigma}
\Big\{M^2_\pi(Q_u+Q_d)T_{1\rho\sigma}+[M^2_\pi Q_u+(2M_K^2-M_\pi^2)Q_s]
T_{2\rho\sigma}\\
&\quad+[M^2_\pi Q_d+(2M_K^2-M_\pi^2)Q_s]T_{3\rho\sigma}
+\frac{1}{3}[M_\pi^2(Q_u+Q_d)+4Q_s(2M_K^2-M_\pi^2)]T_{4\rho\sigma}\\
&\quad
+\frac{2}{3}[M_\pi^2(Q_u+Q_d)+(2M_K^2-M_\pi^2)Q_s]T_{5\rho\sigma}\\
&\quad+M_\pi^2(Q_u-Q_d)\left(\frac{1}{\sqrt{3}}T_{6\rho\sigma}
+\sqrt{\frac{2}{3}}T_{7\rho\sigma}
+\frac{1}{\sqrt{3}}T_{8\rho\sigma}
+\sqrt{\frac{2}{3}}T_{9\rho\sigma}\right)\\
&\quad+\frac{\sqrt{2}}{3}[M_\pi^2(Q_u+Q_d)-2(2M_K^2-M_\pi^2)Q_s]
(T_{10\rho\sigma}+T_{11\rho\sigma})
\Big\}\\
&\quad+(2c_6-c_7+c_8) E^{\rho\sigma}
\Big\{M^2_\pi(Q_u+Q_d)T_{1\rho\sigma}+[(2M_K^2-M_\pi^2)Q_u+M^2_\pi Q_s]
T_{2\rho\sigma}\\
&\quad+[(2M_K^2-M_\pi^2)Q_d+M^2_\pi Q_s]T_{3\rho\sigma}
+\frac{1}{3}[M_\pi^2(Q_u+Q_d)+4Q_s(2M_K^2-M_\pi^2)]T_{4\rho\sigma}\\
&\quad
+\frac{2}{3}[M_\pi^2(Q_u+Q_d)+(2M_K^2-M_\pi^2)Q_s]T_{5\rho\sigma}\\
&\quad+M_\pi^2(Q_u-Q_d)\left(\frac{1}{\sqrt{3}}T_{6\rho\sigma}
+\sqrt{\frac{2}{3}}T_{7\rho\sigma}
+\frac{1}{\sqrt{3}}T_{8\rho\sigma}
+\sqrt{\frac{2}{3}}T_{9\rho\sigma}\right)\\
&\quad+\frac{\sqrt{2}}{3}[M_\pi^2(Q_u+Q_d)-2(2M_K^2-M_\pi^2)Q_s]
(T_{10\rho\sigma}+T_{11\rho\sigma})
\Big\}.
\end{align*}
   Using $2M_K^2-M_\pi^2=M_\pi^2+2(M_K^2-M_\pi^2)$ and noting that $Q_d=Q_s$,
we may write
\begin{align}
\label{LagrangianNLOPVgammaexp2app}
{\cal L}^{VP\gamma}_{\text{NLO,$\chi$}}&=
2M_\pi^2(c_5+c_6-c_7)E^{\rho\sigma}T_{\rho\sigma}\nonumber\\
&\quad+4(M_K^2-M_\pi^2)(c_5+c_6-c_7)E^{\rho\sigma}\Big[
\frac{1}{2}(Q_u+Q_s)T_{2\rho\sigma}+Q_sT_{3\rho\sigma}
+\frac{4}{3}Q_sT_{4\rho\sigma}
+\frac{2}{3}Q_sT_{5\rho\sigma}\nonumber\\
&\quad-2\frac{\sqrt{2}}{3}Q_s(T_{10\rho\sigma}+T_{11\rho\sigma})\Big]\nonumber\\
&\quad+2(M_K^2-M_\pi^2)(c_5-c_6-c_8)E^{\rho\sigma}(Q_s-Q_u)T_{2\rho\sigma}.
\end{align}
   In Table \ref{table_Tifull}, we collect the amplitudes ${\cal A}_i$, $i=1,
   \ldots, 11$, of Eq.~(\ref{leffVPgammaappendix}).
   We have defined $c_+=c_5+c_6-c_7$, $c_-=c_5-c_6-c_8$, and $\tilde{c}_1=c_1+2M_\pi^2c_+$.
   The results depend on 5 parameters $c_1$ ($\tilde{c}_1$), $c_2$, $c_3$, 
$c_+$, and $c_-$.
\renewcommand{\arraystretch}{1.5}
\begin{table}[h]
\caption{Amplitudes ${\cal A}_i$ for the full result in units of $2/F$; 
$c_+=c_5+c_6-c_7$, $c_-=c_5-c_6-c_8$, $\tilde{c}_1=c_1+2M_\pi^2c_+$.}
\label{table_Tifull}
\begin{center}
\begin{tabular}{l l l}
\hline
\hline
Transition \quad & \quad structure \quad & \quad amplitude ${\cal A}_i$ in 
units of $2/F$\\
\hline
$\rho\to\pi\gamma$ \quad & \quad $T_1$ \quad & \quad $\frac{1}{3}\tilde{c}_1$\\
$K^{\ast\pm}\to K^\pm\gamma$ \quad & \quad $T_2$ \quad & \quad $\frac{1}{3}\tilde{c}_1+\frac{2}{3}(M_K^2-M_\pi^2)c_+-2(M_K^2-M_\pi^2)c_-$\\
$K^{\ast}\to K^0\gamma$ \quad & \quad $T_3$ \quad & \quad $-\frac{2}{3}\tilde{c}_1-\frac{4}{3}(M_K^2-M_\pi^2)c_+$\\
$\omega_8\to\eta_8\gamma$\quad & \quad $T_4$\quad & \quad $-\frac{1}{3}\tilde{c}_1-\frac{16}{9}(M_K^2-M_\pi^2)c_+$\\
$\omega_1\to\eta_1\gamma$ \quad & \quad $T_5$ \quad 
& \quad $-\frac{8}{9}(M_K^2-M_\pi^2)c_+$\\
$\rho^0\to\eta_8\gamma$ \quad & \quad $T_6$ \quad & \quad $\frac{1}{\sqrt{3}}\tilde{c}_1$\\
$\rho^0\to\eta_1\gamma$ \quad & \quad $T_7$ \quad & \quad $\sqrt{\frac{2}{3}}\,\tilde{c}_1+\sqrt{\frac{3}{2}}\,c_3$\\
$\omega_8\to\pi^0\gamma$ \quad & \quad $T_8$ \quad 
& \quad $\frac{1}{\sqrt{3}}\tilde{c}_1$\\
$\omega_1\to\pi^0\gamma$ \quad & \quad $T_9$ \quad & \quad $\sqrt{\frac{2}{3}}\,\tilde{c}_1+\sqrt{\frac{3}{2}}\,c_2$\\
$\omega_8\to\eta_1\gamma$ \quad & \quad $T_{10}$ \quad & \quad $\frac{\sqrt{2}}{3}\tilde{c}_1+\frac{1}{\sqrt{2}}c_3
+\frac{8\sqrt{2}}{9}(M_K^2-M_\pi^2)c_+$\\
$\omega_1\to\eta_8\gamma$ \quad & \quad $T_{11}$ \quad & \quad $\frac{\sqrt{2}}{3}\tilde{c}_1+\frac{1}{\sqrt{2}}c_2
+\frac{8\sqrt{2}}{9}(M_K^2-M_\pi^2)c_+$\\
\hline
\hline
\end{tabular}
\end{center}
\end{table}
\renewcommand{\arraystretch}{1}

   In Table \ref{table_TiLNc}, we collect the corresponding coefficients of
the large-$N_c$ expansion.
   We show the results at leading order (LO) and at next-to-leading order 
(NLO), depending on one parameter and six parameters, respectively.
\renewcommand{\arraystretch}{1.5}
\begin{table}[ht]
\caption{Amplitudes ${\cal A}_i$ in the large-$N_c$ expansion in units of $2/F$; $c_+=c_5+c_6-c_7$, $c_-=c_5-c_6-c_8$.}
\label{table_TiLNc}
\begin{center}
\begin{tabular}{l l l}
\hline
\hline
Structure\quad&\quad amplitude ${\cal A}_i$ at  LO in $[2/F]$ \quad&\quad 
amplitude ${\cal A}_i$ at NLO in $[2/F]$\\
\hline
$T_1$\quad&\quad 0 \quad&\quad $\frac{1}{3}c_1-\frac{1}{2}c_4$\\
$T_2$\quad&\quad 0 \quad&\quad $\frac{1}{3}c_1-\frac{1}{2}c_4
-2(M_K^2-M_\pi^2)c_-$\\
$T_3$\quad&\quad $-c_1$ \quad&\quad $-\frac{2}{3}c_1-\frac{1}{2}c_4-2M_K^2c_+$\\
$T_4$\quad&\quad $-\frac{2}{3}c_1$ \quad&\quad $-\frac{1}{3}c_1-\frac{1}{2}c_4-\frac{4}{3}(2M_K^2-M_\pi^2)c_+$\\
$T_5$\quad&\quad $-\frac{1}{3}c_1$ \quad&\quad $-\frac{1}{2}c_2
-\frac{1}{2}c_3-\frac{1}{2}c_4 -\frac{2}{3}(2M_K^2-M_\pi^2)c_+$\\
$T_6$\quad&\quad$\frac{1}{\sqrt{3}}c_1$ \quad&\quad $\frac{1}{\sqrt{3}}c_1+\frac{2}{\sqrt{3}}M_\pi^2 c_+$\\
$T_7$\quad&\quad$\sqrt{\frac{2}{3}}\,c_1$ \quad&\quad $\sqrt{\frac{2}{3}}\,c_1+\sqrt{\frac{3}{2}}\,c_3 
+2\sqrt{\frac{2}{3}}\,M_\pi^2c_+$\\
$T_8$\quad&\quad$\frac{1}{\sqrt{3}}c_1$ \quad&\quad $\frac{1}{\sqrt{3}}c_1+\frac{2}{\sqrt{3}}M_\pi^2 c_+$\\
$T_9$\quad&\quad$\sqrt{\frac{2}{3}}\,c_1$ \quad&\quad $\sqrt{\frac{2}{3}}\,c_1+\sqrt{\frac{3}{2}}\,c_2
+2\sqrt{\frac{2}{3}}M_\pi^2 c_+$\\
$T_{10}$\quad&\quad$\frac{\sqrt{2}}{3}c_1$ \quad&\quad $\frac{\sqrt{2}}{3}c_1+\frac{1}{\sqrt{2}}c_3
+\frac{2}{3}\sqrt{2}(2M_K^2-M_\pi^2)c_+$\\
$T_{11}$\quad&\quad$\frac{\sqrt{2}}{3}c_1$ \quad&\quad $\frac{\sqrt{2}}{3}c_1+\frac{1}{\sqrt{2}}c_2
+\frac{2}{3}\sqrt{2}(2M_K^2-M_\pi^2)c_+$\\
\hline
\hline
\end{tabular}
\end{center}
\end{table}
\renewcommand{\arraystretch}{1}

\end{appendix}

\end{document}